\documentclass[preprints,article,accept,moreauthors,pdftex,a4paper]{Definitions/mdpi}

\pdfoutput=1

\firstpage{1} 
\pubvolume{13}
\issuenum{9}
\articlenumber{1655}
\pubyear{2021}
\copyrightyear{2021}
\history{Received: date; Accepted: date; Published: date}

\usepackage{textcomp} 
\usepackage{gensymb} 
\usepackage{bm} 

\newcommand\MSG[1]{\mbox{MSG-#1}} 

\Title{Dual View on Clear-Sky Top-of-Atmosphere Albedos from Meteosat Second Generation Satellites}

\Author{Alexandre Payez *\orcidA{}, Steven Dewitte\orcidB{} and Nicolas Clerbaux\orcidC{}}

\AuthorNames{Alexandre Payez, Steven Dewitte and Nicolas Clerbaux}

\address[1]{%
Royal Meteorological Institute of Belgium, Ringlaan 3 Avenue Circulaire, B-1180 Brussels, Belgium; alexandre.payez@meteo.be (A.P.); steven.dewitte@meteo.be (S.D.); nicolas.clerbaux@meteo.be (N.C.)
}

\corres{Correspondence: alexandre.payez@meteo.be}

\abstract{%
Geostationary observations offer the unique opportunity to resolve the diurnal cycle of the Earth's Radiation Budget at the top of the atmosphere (TOA), crucial for climate-change studies. 
However, a drawback of the continuous temporal coverage of the geostationary orbit is the fixed viewing geometry. 
As a consequence, 
imperfections in the angular distribution models (ADMs) used in the radiance-to-flux conversion process or residual angular-dependent narrowband-to-broadband conversion errors can result in systematic errors of the estimated radiative fluxes.
In this work, focusing on clear-sky reflected TOA observations, we compare the overlapping views from Meteosat Second Generation satellites at $0\degree$ and $41.5\degree$E longitude which enable a 
quantification of viewing-angle-dependent differences.
Using data derived from the Spinning Enhanced Visible and InfraRed Imager (SEVIRI), we identify some of the main sources of discrepancies, and show that they can be significantly reduced at the level of one month.
This is achieved, separately for each satellite, via a masking procedure followed by an empirical fit at the pixel-level that takes into account all the clear-sky data from that satellite, calculated separately per timeslot of the day, over the month of November 2016.
The method is then applied to each month of 2017, and gives a quadratic mean of the albedo root-mean squared difference over the dual-view region which is comparable from month to month, with a 2017 average value of 0.01.
Sources of discrepancies include the difficulty to estimate the flux over the sunglint ocean region close to the limbs, 
the fact that the data processing does not include dedicated angular distribution models for the aerosol-over-ocean case, and the existence of an observer-dependent diurnal-asymmetry artefact affecting the clear-sky-albedo dependence on the solar zenith angle particularly over land areas.}


\begin{document}

\section{Introduction}


The Earth's Radiation Budget (ERB) at the top of the atmosphere (TOA) is a crucial observable and one of the Essential Climate Variables~\cite{ECV} defined by the Global Climate Observing System (GCOS) for climate-change studies.
For the Earth system, it describes the energy balance between what comes in from the Sun and what leaves the Earth, both as reflected solar (shortwave) radiation and as outgoing thermal (longwave) emission.
The global warming currently ongoing with unprecedented speed since the mid-20th century is a direct consequence of an overall imbalance in the ERB, predominantly caused by human activities from the industrial revolution to the present~\cite{IPCC_AR5:2014}. 

Characterising as best as possible the state of the ERB can only be done from space, and for that we crucially need observer-independent quantities to be retrieved, knowing that satellite instruments can only directly provide observer-dependent radiances $L$ in a given direction at a given time.\footnote{Note that wide-field-of-view instruments, which cannot provide spatial information, are not considered in this work.}
Going from the measured observer-dependent radiances to such observer-independent quantities involves two important steps. The first is called unfiltering and compensates for the spectral-dependence of the engineered instrument~\cite{Green_etal:1996}. The second is the inversion\hspace{1pt}---\hspace{1pt}also known as angular conversion\hspace{1pt}---\hspace{1pt}which, from that unfiltered radiance in a given direction then associates an estimated observer-independent irradiance appropriate for the observed scene~\cite{Suttles_etal:1992}, typically via a collection of empirical angular distribution models (ADMs), such as the Clouds and the Earth's radiant Energy System (CERES) Tropical Rainfall Measurement Mission (TRMM) ADMs~\cite{ADM_CERES_TRMM:2003}.
Formally, the irradiance or `flux' $F(\theta_\odot)$ leaving an imaginary surface element at the top of the atmosphere (in units of W~m$^{-2}$) is the integral, over all the outgoing solid angles in a hemisphere, of the individual radiances $L(\theta_\odot, \theta_{\rm vz}, \phi_{\rm rel})$ (in units of W~m$^{-2}$~sr$^{-1}$) leaving that surface element~\cite{Suttles_etal:1988}:
\begin{align}
	\label{eq:flux}
	F(\theta_\odot) &= \int_{\mathrm{hemisphere}^{\uparrow}} L(\theta_\odot, \theta_{\rm vz}, \phi_{\rm rel}) \cos\theta_{\rm vz} \ d\Omega \\
	  &= \int_{0}^{2\pi} d\phi_{\rm rel} \int_{0}^{\frac{\pi}{2}} L(\theta_\odot, \theta_{\rm vz}, \phi_{\rm rel}) \cos\theta_{\rm vz} \sin\theta_{\rm vz} \ d\theta_{\rm vz}, 
\end{align}
where $\theta_\odot$ is the solar zenith angle, $\theta_{\rm vz}$ is the viewing zenith angle, and $\phi_{\rm rel}$ is the relative azimuth angle.
As well-known, this relation simply reduces to $F(\theta_\odot$) = $\pi L(\theta_\odot)$ when the flux is isotropic (Lambertian).
ADMs are then typically introduced at this point~\cite{Suttles_etal:1988}, and provide anisotropic factors which compare an equivalent Lambertian flux $\pi L(\theta_\odot, \theta_{\rm vz}, \phi_{\rm rel})$ to the actual flux $F(\theta_\odot)$ given in Eq.~\eqref{eq:flux}:

\begin{equation}
	R(\theta_\odot, \theta_{\rm vz}, \phi_{\rm rel}) = \frac{\pi L(\theta_\odot, \theta_{\rm vz}, \phi_{\rm rel})}{F(\theta_\odot)}.
	\label{eq:ADMs}
\end{equation}
With a proper set of empirical ADMs, Eq.~\eqref{eq:ADMs} can then be inverted to derive the flux from the radiance.
Finally, the albedo 
(a dimensionless quantity taking values between 0 and 1) is defined as the ratio of the reflected and incoming solar fluxes:

\begin{equation}
	a(\theta_\odot) = \frac{F_{\rm reflected}(\theta_\odot)}{F_{\rm incoming}(\theta_\odot)}, \qquad \textrm{where } F_{\rm incoming}(\theta_\odot) = E_\odot \cos\theta_\odot;
	\label{eq:albedo}
\end{equation}
$E_\odot$ being the solar constant \cite{Dewitte_Nevens:2016} corrected for the Sun--Earth distance~\cite{Bretagnon_Francou:1988}.

For climate models, it is moreover important to precisely monitor the internal structures of the ERB over the globe and throughout the day, seeing that these govern important aspects of the climate on our planet; in particular, its diurnal cycle, and especially the formation of clouds during the day, is a key component of the tropical climate \cite{Slingo}.
Most dedicated ERB observations thus far have been made from Low-Earth-Orbit (LEO) satellites.
Chiefly among these are Sun-synchronous polar orbiters, such as NASA's Terra and Aqua satellites, which have been observing over the last two decades with the CERES instrument~\cite{Wielicki_etal:1996} each location on Earth\hspace{1pt}---\hspace{1pt}though not more than twice per day outside of the polar regions.
The joint EUMETSAT/ESA geostationary (GEO) Meteosat Second Generation (MSG) satellites are actually in a unique position to complement these observations and observe the diurnal cycle, and this for two reasons. 
The first is that their geostationary orbits offer the advantage of an excellent temporal sampling of the observed locations throughout the day. 
The second is that they embark both the multispectral Spinning Enhanced Visible and InfraRed Imager (SEVIRI) with 12 spectral channels at a nadir resolution of 3~km (1~km for the high-resolution visible channel)~\cite{MSG}, and the broadband Geostationary Earth Radiation Budget (GERB) instrument with nadir resolution of 50~km~\cite{Harries_etal:2005}.
Every 15 minutes both these instruments provide observations of the full Earth disk seen from the satellite viewpoint.
They do have the disadvantage of fixed viewing angles for each pixel location in the Meteosat domain though. Compared to LEO satellites, this makes the measurements of the broadband TOA radiative fluxes more sensitive to angular-dependent errors.
It is therefore particularly important to address any remaining observer-dependent systematics in flux or albedo data, knowing that such unphysical artefacts would introduce errors, \textit{e.g.\@} in the Earth's Radiation Budget determination. 

Interestingly, we now actually have the opportunity to cross-check retrieved GEO products. Indeed, since the end of the year 2016, there have been two MSG satellites at different longitudes and with overlapping scenes.
One of them is at 0$\degree$ longitude (\MSG{3}, replaced by \MSG{4} on 2018/02/20), and the other is at 41.5$\degree$E longitude, providing the 'Indian Ocean Data Coverage' (\MSG{1}, which should be replaced by \MSG{2} at 45.5$\degree$E longitude in 2022; see \textit{e.g.\@} Ref.~\cite{Ratier_presentation:2020}).
Taking advantage of this dual view, our aim with this work is then to compare, quantify the discrepancies and try to address them in order to make GEO-derived products as useful as possible.
For simplicity the focus will only be on the clear-sky top-of-atmosphere broadband shortwave albedo, and we will often simply refer to it as the `albedo'.
In the following, we are only going to use and consider pure SEVIRI synthetic products: the so-called `GERB-like' GL-SEV products~\cite{Dewitte:2008}.\footnote{We use what are, at the time of writing, the very latest available products covering the dual-view period: GL-SEV HR V003. These can be obtained via the \href{https://gerb.oma.be}{https://gerb.oma.be} website.}
Although these products are derived in the context of the GERB project, note that they do not use any GERB radiances. Involving a narrowband-to-broadband procedure~\cite{Clerbaux_etal:2008} and relying on the CERES TRMM ADMs~\cite{ADM_CERES_TRMM:2003} for the angular-conversion process, these products consist of HDF5 files available every 15 minutes (96 per day), containing $3\times3$~SEVIRI-pixel image layers (9~km $\times$ 9~km resolution at nadir). 
The CERES TRMM ADMs were selected by the GERB team since they were available at the time of the GERB-processing development, and especially since they cover all angular configurations (TRMM was a low-inclination LEO satellite, eventually covering all solar zenith angles as its orbit was precessing). A drawback of these ADMs is that they do not include dedicated ADMs for the case of aerosols over the ocean. That case is actually covered in a more recent version of the CERES ADMs, derived from the Terra and Aqua polar-orbiting satellites~\cite{Su_etal:2015}.

This paper is organised as follows. In Section~\ref{sec:comparison_method}, we present the basic methodology that we use to compare the results from the different satellites.
We then show in Section~\ref{sec:one_month} how the monthly consistency between the two views can be improved for the month of November 2016, via masks and an empirical-fit procedure. 
Finally, Section~\ref{sec:one_year} shows that similar results can be obtained for the entire year 2017, before we conclude in Section~\ref{sec:conclusions}.

\section{Dual-view comparison method}
\label{sec:comparison_method}

The same treatment is applied independently to GL-SEV images from \MSG{3} (SEV3) and \MSG{1} (SEV1), and is done one month at a time.
For that month, and for each individual HHMM timeslot among the 15-minute timeslots, our basic treatment is the following:
\begin{itemize}
	\item for each day at that timeslot, we first derive the instantaneous clear-sky TOA albedo image after applying the appropriate masks\hspace{1pt}---\hspace{1pt}\textit{e.g.\@} keeping only those pixels for which the cloud-cover layer is zero (as discussed in the following, we are going to apply a number of extra conditions);
	\item then, similarly to what is done for monthly hourly products\footnote{For monthly hourly products, see for instance the CERES SYN1deg~\cite{Doelling_etal:2016} and the CM~SAF MMDC~\cite{Urbain_etal:2017} and~\cite{Clerbaux_etal_DATA:2017} products.}, we use all these images to calculate the monthly “representative albedo image” at that timeslot: in this work, we considered both the mean and the median, and unless otherwise stated we will show results obtained using the median albedo (robust statistics);
	\item in a common $0.5\degree \times 0.5\degree$ latitude--longitude grid, we can then calculate the grid-box difference for that specific timeslot; in other words, each pixel location $x$ will store the albedo difference
	\begin{equation}
		x = x_{\rm SEV3} - x_{\rm SEV1},
	\end{equation}
	which would be 0 for all pixel if the calibration of SEV1 and SEV3 was the same, if there were no narrowband-to-broadband nor angular-dependent errors, and if the impact of having different clear-sky atmospheric paths from a given scene to each of the two satellites can be reduced to a strictly angular issue\hspace{1pt}---\hspace{1pt}an assumption already implicit in the radiance-to-flux conversion.
\end{itemize}

Once this treatment is done, we can proceed and apply statistics over the whole set of 96 \mbox{`${\rm SEV3} - {\rm SEV1}$'} images: for each pixel $x$, using all of the $(t=1, ..., n_{t_x})$ timeslots for which we have non-masked data at that location, we calculate the root-mean-squared difference or root-mean-squared deviation (\textit{i.e.\@} the square root of the mean squared albedo difference):

\begin{equation}
	{\rm RMSD} = \sqrt{  \frac{1}{n_{t_x}} \sum_{t=1}^{n_{t_x}} x_t^2  };\label{eq:RMSD}
\end{equation}
the bias (\textit{i.e.\@} the mean albedo difference):

\begin{equation}
	\mathrm{bias} = \frac{1}{n_{t_x}} \sum_{t=1}^{n_{t_x}} x_t;
\end{equation}
and finally the bias-corrected standard deviation $\sigma$, from the corresponding variance:

\begin{equation}
	\sigma^2 =  \frac{1}{n_{t_x}} \sum_{t=1}^{n_{t_x}} (x_t - \mathrm{bias})^2 = {\rm RMSD}^2 - {\rm bias}^2.
\end{equation}

\section{One month: sample results and improvements}
\label{sec:one_month}

\subsection{Raw GL-SEV products}

Since our aim is to assess and compare the retrieved products from the two different MSG satellites, here we use the fluxes exactly as they are given in the GL-SEV products. Note in particular that, as part of the data processing of these products, a shortwave flux over the sunglint region is provided~\cite{Bertrand_etal:2006_sunglint}, and that the case of `aerosol over ocean' pixels is treated there as `clear-sky ocean' with the use of identical CERES TRMM ADMs.

\begin{figure}[ht]
\centering

	\includegraphics[height=5.079375cm, trim = 0 24 54 0, clip]{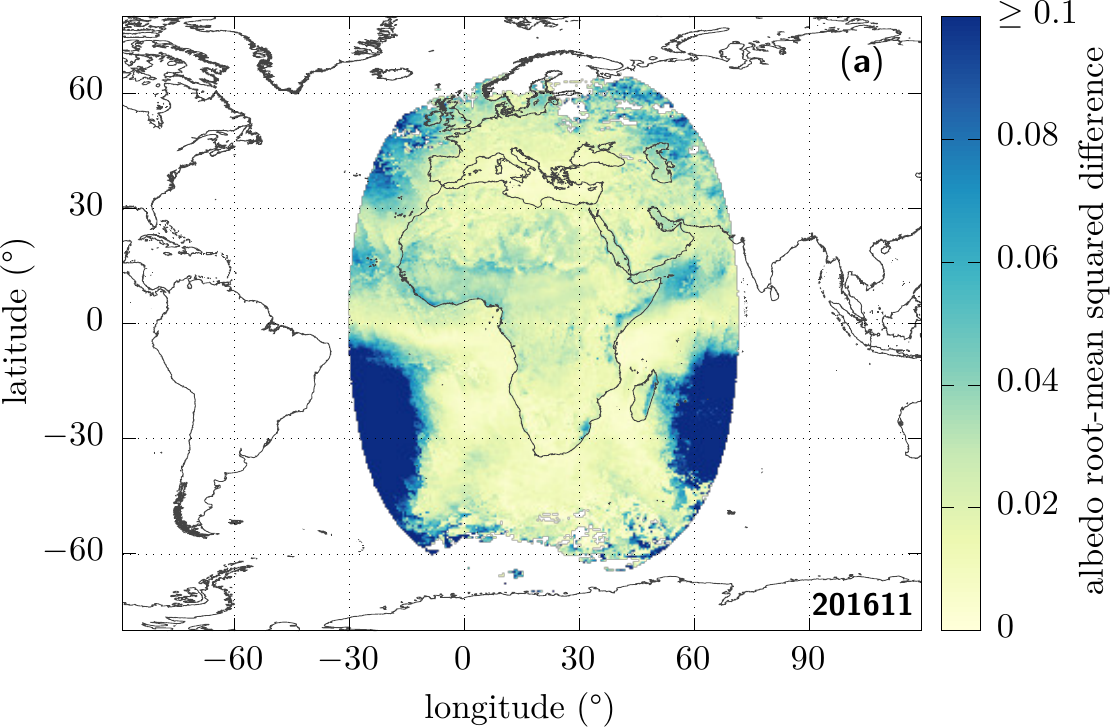}
	\includegraphics[height=5.079375cm, trim = 35 24 0 0, clip]{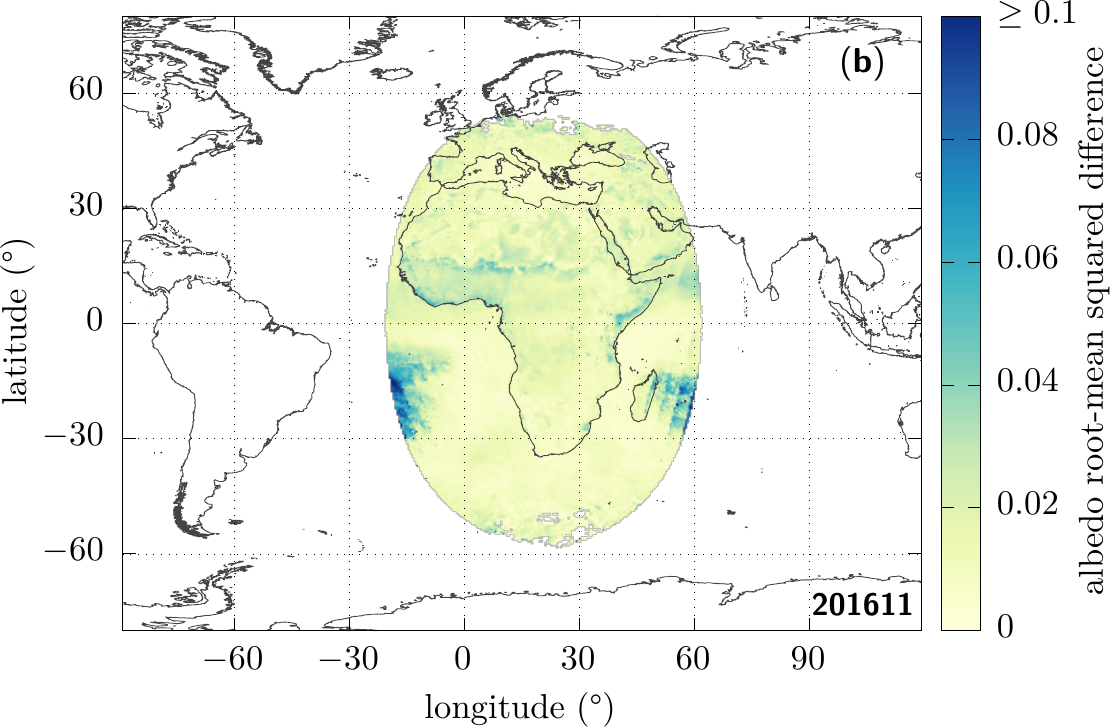} 
	\includegraphics[height=5.7375cm, trim = 0 0 54 0, clip]{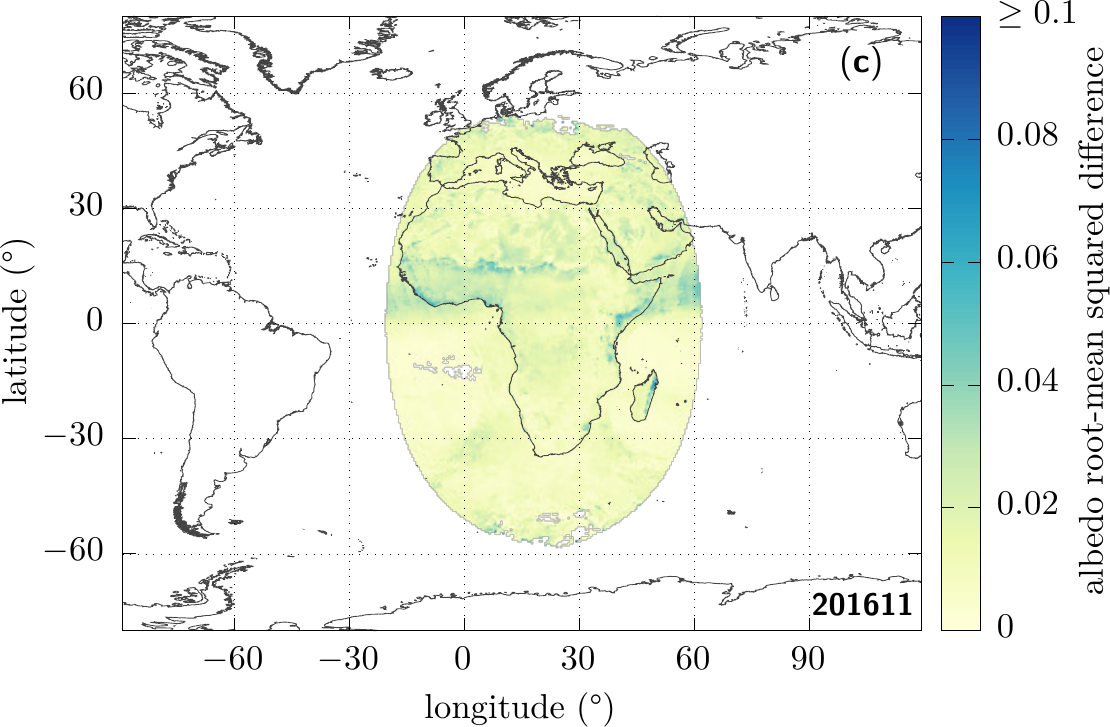} 
	\includegraphics[height=5.7375cm, trim = 35 0 0 0, clip]{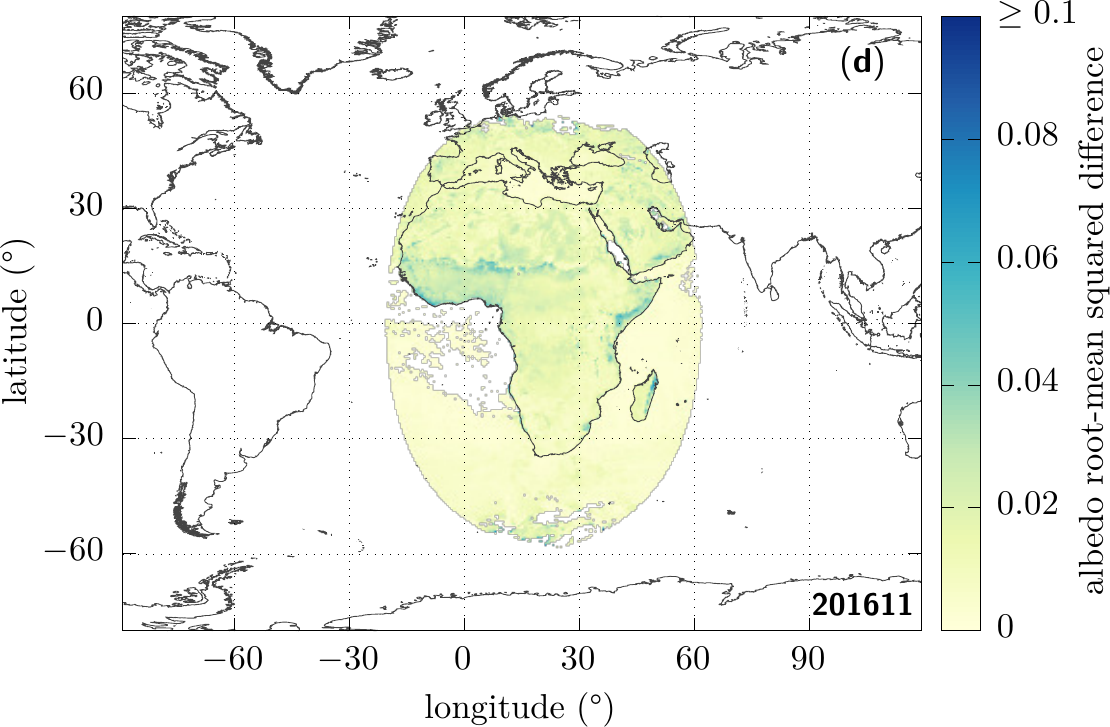}
\caption{Albedo root-mean squared difference (RMSD, Eq.~\eqref{eq:RMSD}) over the dual-view region seen by both \MSG{3} and \MSG{1} for the month of November 2016, after applying different masks. (\textbf{a}, \emph{top left}) Clear-sky, without further constraints: the shortwave flux information is used wherever it is defined, simply provided that the corresponding cloud cover at that time and place is zero. (\textbf{b}, \emph{top right}) Same as (\textbf{a}), but only if the solar and the viewing zenith angles are $<70\degree$. (\textbf{c}, \emph{bottom left}) Same as (\textbf{b}), but now adding a sunglint mask, requiring the 
sunglint angle to be $>25\degree$ over ocean pixels. (\textbf{d}, \emph{bottom right}) Same as (\textbf{c}), now adding an aerosol mask over ocean, requiring that the aerosol optical depth at 0.6~$\mu$m is $<0.1$, and extending the sunglint mask to retain only 
sunglint angles $>40\degree$ over ocean pixels.}
\label{fig:latlon_orig}
\end{figure}

Figure~\ref{fig:latlon_orig} (\textbf{a}) shows the root-mean-squared-difference result that we obtain from our comparison when we naively use the shortwave fluxes whenever they are defined in the GL-SEV products.\footnote{The albedo is set to 1 where it exceeds this value. Such pixels are not masked, to help identify and address potential issues.} Among the most salient features that we can see is that there are issues related to unidentified clouds, which actually occur mainly at large viewing and solar zenith angles. What is then probably the most striking difference are the large extended regions (dark blue) in the bottom left and the bottom right of the image, where the discrepancies can be larger than 0.10. These can actually be linked to the sunglint regions (when the viewing direction coincides with the specular sunlight reflection off the ocean surface) getting close to either of the limbs of the overlapping area: the left one corresponds to \MSG{1} and the right one, to \MSG{3}. Other areas over the ocean with discrepancies which can reach $\sim 0.05$ are also visible close to the west and east coasts of Africa (blue), and are actually due to the presence of aerosols.
These are serious discrepancies, given how low the albedo over clear-sky ocean typically is ($\sim 0.06$ under overhead solar conditions).
Since the ocean covers about 70\% of the Earth, it is particularly important to correct these. 
Finally, further discrepancies are found over land.

\subsection{Masks}\label{sec:masks}

We can now try to improve the consistency between the albedo values retrieved from the two satellites by first applying a number of masks and thresholds.

First of all, what is frequently done in the literature is to cut on the zenith angles as the quality is expected to suffer from increasing errors at large values; see \textit{e.g.\@} Refs.~\cite{GERB_QualitySummary, Loeb_etal:1999}. The exact threshold may slightly differ from paper to paper; here we use $\theta_\odot, \theta_{\rm vz} < 70\degree$.\footnote{For consistency with an aerosol-optical-depth mask applied in the following, as there is no retrieval at larger zenith angles.}
After doing that, we can see that things are already improved (see Figure~\ref{fig:latlon_orig}~(\textbf{b})), but notice that the discrepancies which we identified for the sunglint and the aerosol areas in particular actually remain of the same order, respectively $>0.10$ and $\sim 0.05$.

Since very large differences coming from the shortwave flux over the sunglint regions remain, we then remove the modelled fluxes provided in those regions by masking all the ocean pixels for which the sunglint angle\footnote{The sunglint angle is defined via: $\cos(\mathrm{sunglint~angle}) = \sin(\theta_\odot) \sin(\theta_{\rm vz}) \cos(\phi_{\rm rel}) + \cos(\theta_\odot) \cos(\theta_{\rm vz})$.} is smaller than 25$\degree$.
Doing so gives Figure~\ref{fig:latlon_orig} (\textbf{c}).

\begin{figure}[h]
\centering

	\includegraphics[height=5.205cm, trim = 0 0 15 0, clip]{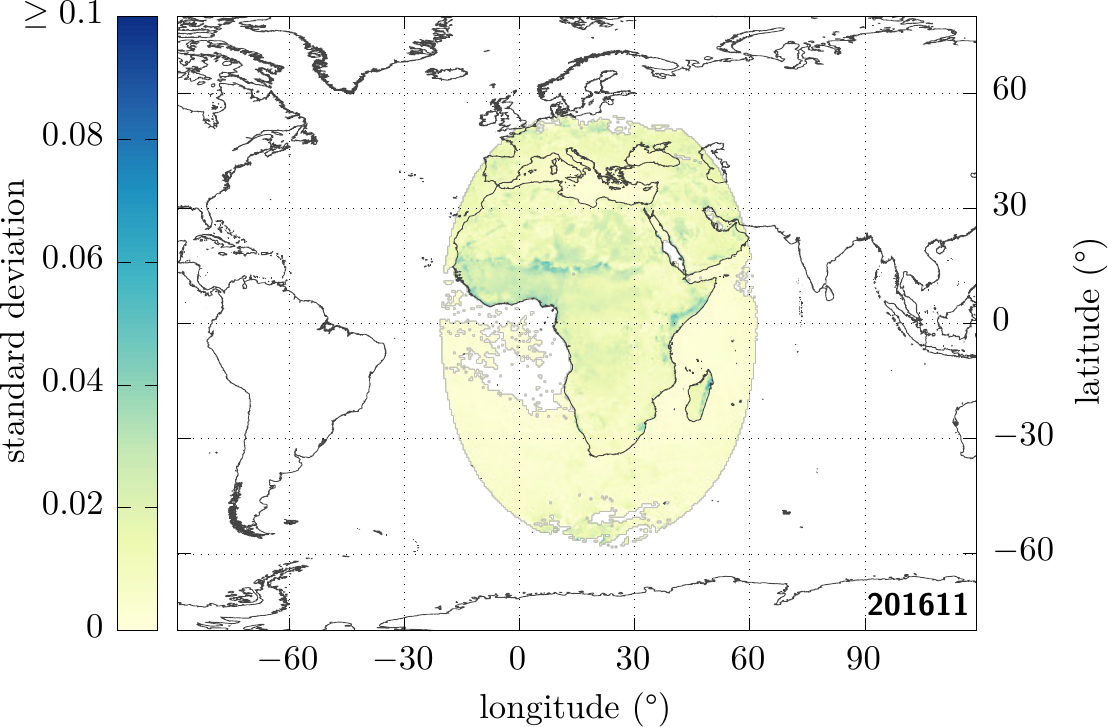}
	\hfill
	\includegraphics[height=5.205cm, trim = 0 0 0 0, clip]{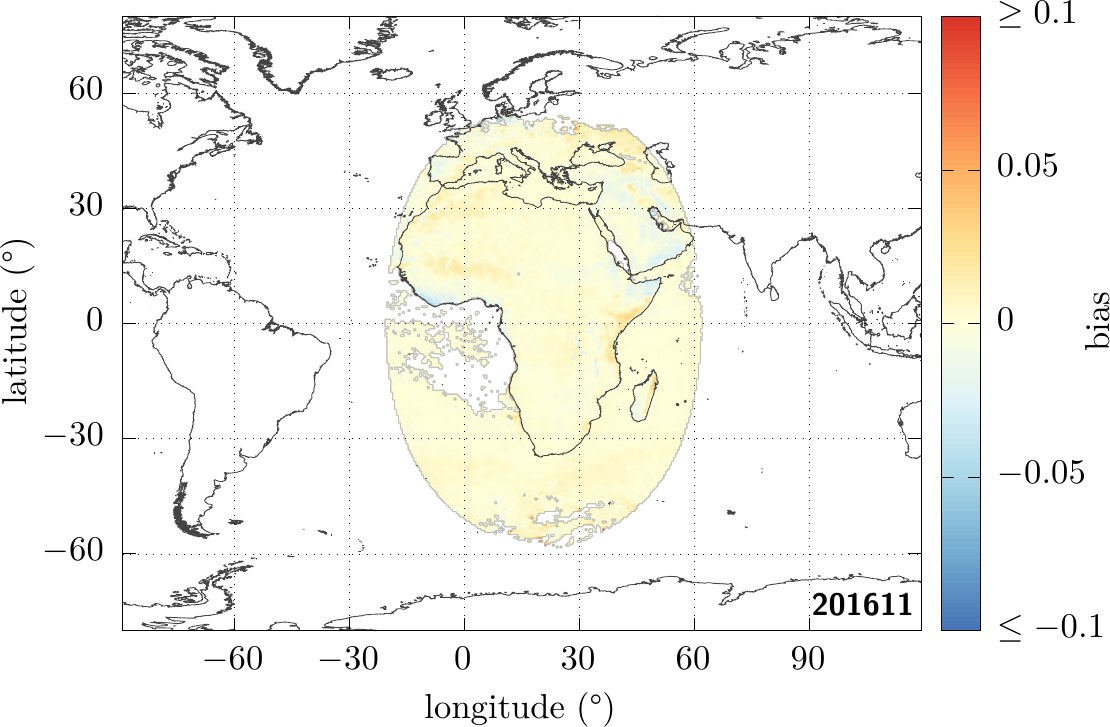}

\caption{Decomposition of Figure~\ref{fig:latlon_orig} (\textbf{d}) in terms of standard deviation (\emph{left}) and bias (\emph{right}).}
\label{fig:latlon_sigmabiasorig}
\end{figure}

After applying these different masks, the discrepancies related to aerosols over the ocean are still clearly visible and still of the order of $\sim 0.05$. Due to the lack of dedicated ADMs for that case\footnote{Note that the CERES team proposes a method to account for aerosols in the clear-sky ocean ADM; see Ref~\cite{ADM_CERES_TRMM:2003}. However, this correction is not actually applied in the GL-SEV processing.}, we are going to mask these aerosol-loaded regions over ocean pixels.\footnote{The presence of aerosols results in a quite diffuse reflection; very different from the strong specular reflection in the clear-sky ocean case. The use of clear-sky ocean ADMs that do not take aerosols into account is therefore particularly problematic.} There is an aerosol-optical-depth retrieval over the ocean in the GERB/GL-SEV processing~\cite{DePaepe_etal:2008}; we can therefore use the aerosol-optical-depth observations at 0.6~$\mu$m, and only keep those pixels for which this optical depth is smaller than 0.1. For consistency with the aerosol-retrieval algorithm and to avoid introducing artefacts in the images, the sunglint mask used in the previous step is then enlarged to only keep 
sunglint angles greater than 40$\degree$ over ocean pixels. This is our final result using masks; it is shown in Figure~\ref{fig:latlon_orig}~(\textbf{d}). Its decomposition in terms of standard deviation and bias is given in Figure~\ref{fig:latlon_sigmabiasorig}, from which one can see that the standard deviation typically contributes much more to the RMSD than does the bias.

Now, if we furthermore compare the statistics over the entire image obtained when naively using the fluxes as provided in the GL-SEV products (Figure~\ref{fig:latlon_orig} (\textbf{a})) to the ones that we obtained after the masking procedures we just described (Figure~\ref{fig:latlon_orig} (\textbf{d})), we obtain the results given in Table~\ref{table:201611_withmasks}.

\begin{table}[H]
\caption{Summary statistics, where $\langle \cdot \rangle$ stands for the average over all the contributing pixels.}
\centering
\begin{tabular}{lccc}
\toprule
&$\bm{\sqrt{\langle\mathrm{RMSD^2}\rangle}}$	&$\bm{\sqrt{\langle\sigma^2\rangle}}$	& $\bm{\langle\mathrm{|bias|}\rangle}$\\
\midrule
naively &	0.055		& 0.052		& 0.012 \\
with all the masks &	0.014		& 0.013		& 0.004 \\
\bottomrule
\end{tabular}
\label{table:201611_withmasks}
\end{table}

These summary statistics provide a single value after looping over all the ($x = 1, \dots, N_x$) non-masked pixels in the corresponding image.
More precisely, denoting the RMSD, $\sigma$, and bias of a given pixel via a $\cdot_x$ subscript, the quantities given in that table are the square root of the average mean squared difference: 

\begin{equation}
	\sqrt{\langle \mathrm{RMSD^2} \rangle} = \sqrt{ \frac{1}{N_x} \sum_{x=1}^{N_x} {\mathrm{RMSD}_x}^2 };
	\label{eq:quadraticmean_RMSD}
\end{equation}
the square root of the average variance:

\begin{equation}
	\sqrt{\langle \sigma^2 \rangle} =  \sqrt{ \frac{1}{N_x} \sum_{x=1}^{N_x} {\sigma_x}^2 } ;
\end{equation}
and finally, the mean absolute bias (or mean absolute difference):

\begin{equation}
	\langle |\mathrm{bias}| \rangle = \frac{1}{N_x} \sum_{x=1}^{N_x} \left| \mathrm{bias}_x \right|.
\end{equation}
In particular, we see in Table~\ref{table:201611_withmasks} that the quadratic mean of the RMSD over all pixels, Eq.~\eqref{eq:quadraticmean_RMSD}, is significantly reduced, going from 0.055 to 0.014.

The consistency is much improved after this masking procedure, but we stress once again that we actually had to mask the aerosol over ocean case in order to obtain these results, and that having access to correct fluxes also in such cases is clearly a high priority; see \textit{e.g.\@} Refs.~\cite{Zhang_etal:2005, Su_etal:2015}.

\subsection{Diurnal-asymmetry artefact} \label{sec:diurnal_asymmetry}

We now move on to discuss a well-known issue with geostationary data: due to the fixed viewing geometry from GEO, there can be a systematic diurnal-asymmetry artefact in clear-sky GEO shortwave fluxes~\cite{Bertrand_etal:2006_asym}; see also Ref.~\cite{EBAF:2018}.

\begin{figure}[ht]
\centering
	\includegraphics[height=5.7375cm]{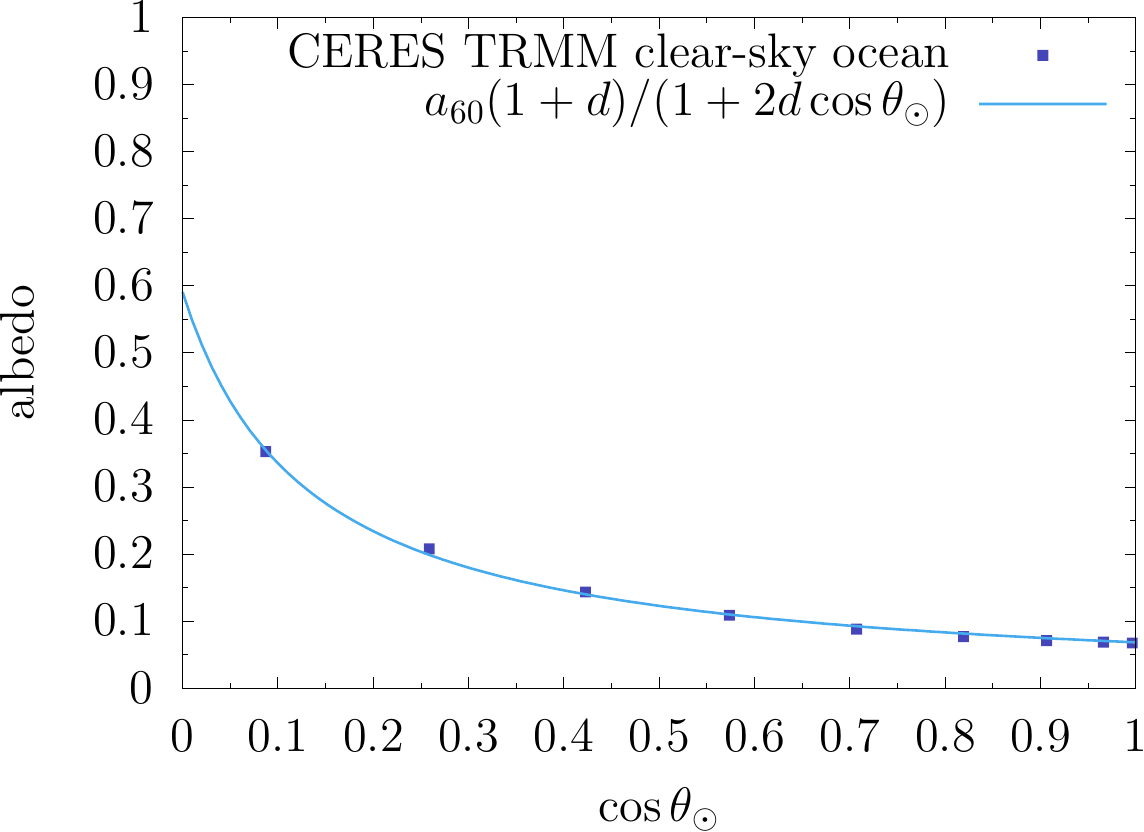}
\caption{CERES TRMM ADMs: albedo in the clear-sky ocean case, fit with the functional form Eq.~\eqref{eq:albedo_functional_form}.}
\label{fig:fit_clearskyocean_ADMs}
\end{figure}

In the majority of cases however, no asymmetry in the clear-sky top-of-atmosphere albedo is actually expected between the (local time) morning and afternoon. That is, at least if the surface properties do not change over time~\cite{Minnis_etal:1997}.
Indeed, the albedo over land essentially evolves as a function of the solar zenith angle, which is symmetrical with respect to the local noon~\cite{Dickinson:1983, Briegleb_etal:1986}:

\begin{equation}
	a(\theta_\odot) = a_{60} \frac{1+d}{1+2d\cos\theta_\odot},
	\label{eq:albedo_functional_form}
\end{equation}
with parameters $a_{60} = a(\theta_\odot = 60\degree{})$ and $d$, a dimensionless parameter tied to how exactly the albedo changes with $\cos\theta_{\odot}$.
Equation~\eqref{eq:albedo_functional_form}
is extensively found in the literature, both for TOA and surface albedo studies as a function of the solar zenith angle; see Ref.~\cite{Yang_etal:2008} for a very detailed study which also reviews the literature. 
This functional form was first introduced in Ref.~\cite{Dickinson:1983} for the case of infinite canopies. It was then used in Ref.~\cite{Briegleb_etal:1986} for all types of land surfaces.
As shown in Figure~\ref{fig:fit_clearskyocean_ADMs}, one can check against the CERES ADMs that the same functional form can be used over ocean pixels (rms of residuals: 0.005); this is using the average wind speed model.

Knowing that the albedo, just like the flux, is expected to be only a function of the solar zenith angle at any single time and knowing the general functional form of this dependence, we can then take benefit of the GEO temporal sampling and consider all the observations at different times in order to obtain a more robust and consistent result at the level of one month. 
Following this method will provide a corrected monthly mean (level-3) GL-SEV product; it does not directly correct the instantaneous level (level-2).
Note that this assumes that the surface properties and the atmospheric transmission affecting the clear-sky top-of-atmosphere radiation do not significantly change over that period of one month.

\begin{figure}[ht]
\centering

	\includegraphics[height=5.7375cm]{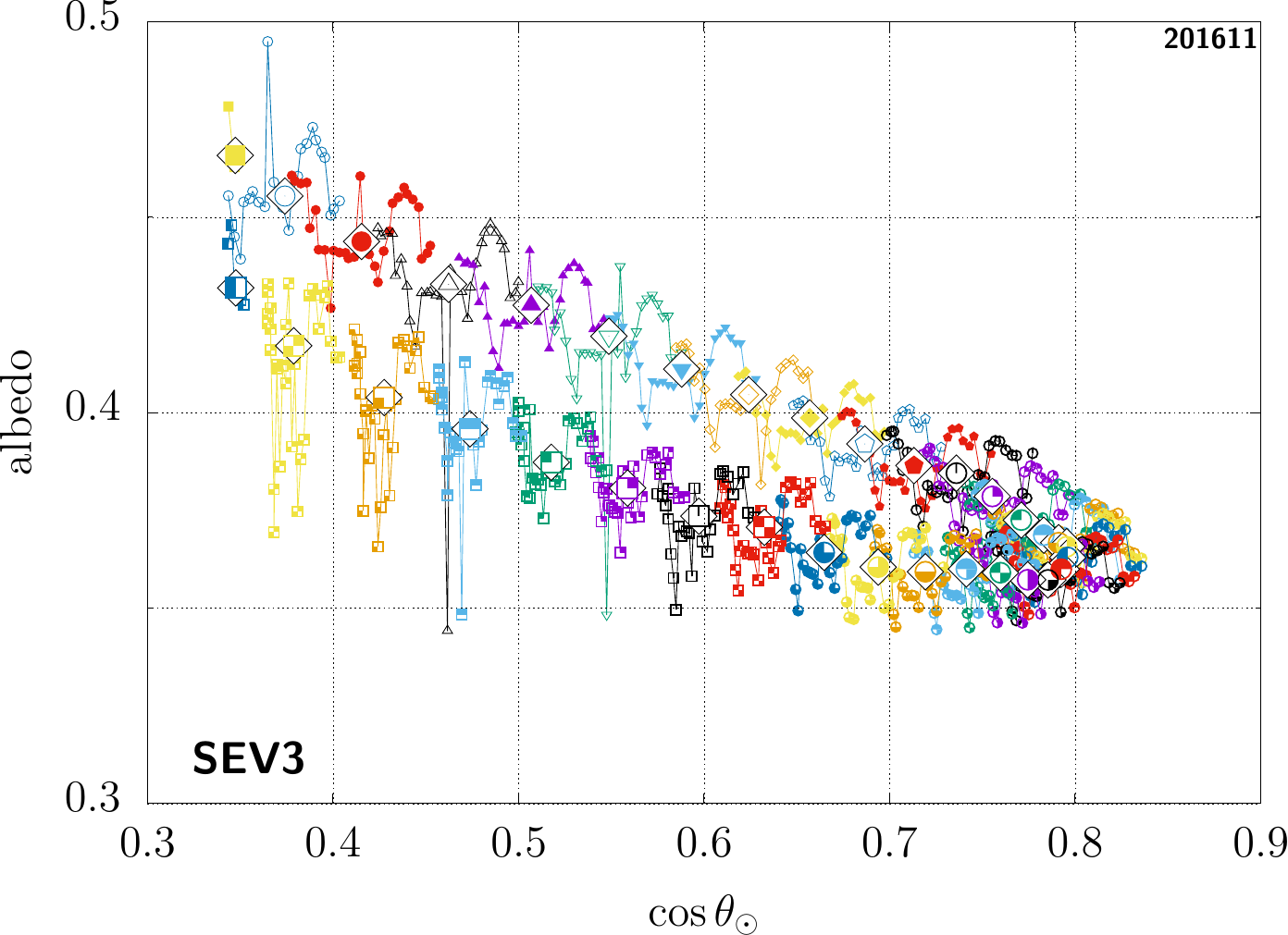}
	\includegraphics[height=5.7375cm, trim = 20 0 0 0, clip]{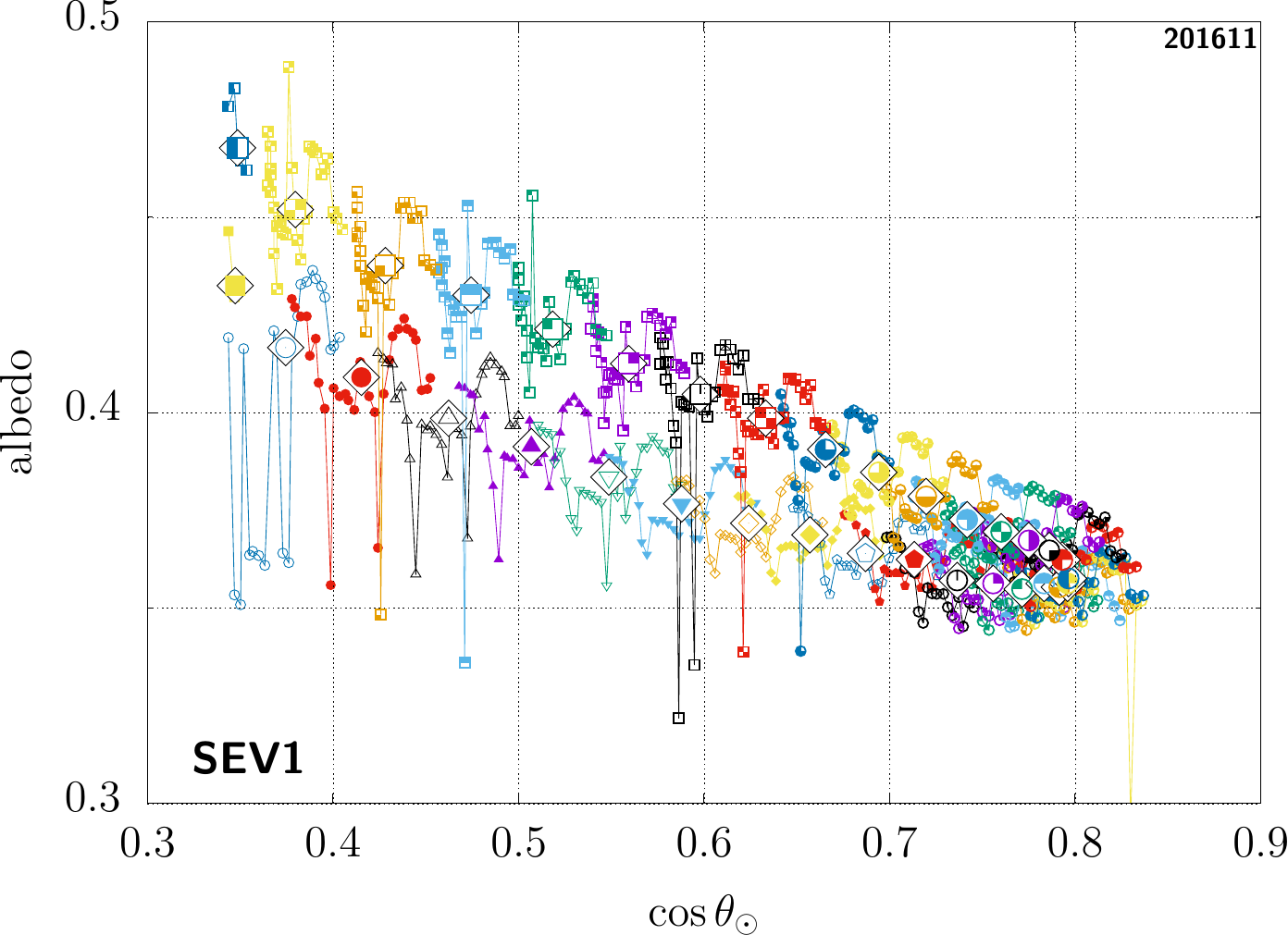}

\caption{Diurnal-asymmetry artefact for the same 
GEO pixel (18.62$\degree$ latitude, 25.52$\degree$ longitude) in the Sahara, seen from \MSG{3} (\emph{left}) and \MSG{1} (\emph{right}); see main text. For each timeslot (\textit{i.e.\@} each colour + symbol pair), the large symbol inside a diamond shape shows the median albedo versus the mean $\cos\theta_{\odot}$ for that timeslot, while the corresponding smaller symbols denote the instantaneous data used to compute those.}%
\label{fig:samplepixel_sahara}
\end{figure}

In Figure~\ref{fig:samplepixel_sahara} (\emph{left}), we show an example of the diurnal-asymmetry artefact in a case of a rarely cloudy sample pixel taken in the Sahara, for which we can therefore get a very detailed view of the clear-sky-TOA-albedo temporal evolution throughout the day. This is seen from \MSG{3} and we can clearly see that there are in fact two branches for similar values of the cosine of the solar zenith angle, the higher branch in this case corresponding to the morning, and the afternoon branch being the lower one.
In this figure, each pair of colour and small symbols corresponds to the instantaneous clear-sky-TOA albedo for different days at the same timeslot. For each of these, we also overlay a summary representative (mean or median) value for that timeslot as a larger symbol inside a diamond shape\hspace{1pt}---\hspace{1pt}unless otherwise stated all such results shown here were obtained using the median albedo versus the mean cosine of the solar zenith angle at the corresponding timeslot.

In Figure~\ref{fig:samplepixel_sahara} (\emph{right}), we show the same case, but now seen from \MSG{1}. Notice how the two branches are then actually swapped: the higher branch corresponding to the afternoon in this case, and the morning branch being the lower one. This complete contradiction at the qualitative level highlights the fact that this asymmetry cannot be of physical origin.\footnote{While the branches themselves cannot be physical, one can notice the presence of overimposed instantaneous patterns.
In contrast to the branches, those are actually consistent between the two satellites at corresponding times and likely of physical origin. They are in fact already present in SEVIRI narrowband radiances. See Appendix~\ref{app:patterns}.} The problem is rather related to imperfections in the ADMs used to convert radiances into flux or in the narrowband-to-broadband procedure, which, due to the fixed view of GEO observations, are then turned into systematic errors.\footnote{Although an effect due to residual narrowband-to-broadband errors in GL-SEV cannot be excluded, this would not be sufficient to explain the diurnal-asymmetry artefact, which also appears in pure broadband GERB data; see Appendix~\ref{app:gerb}.} This well-known situation is not specific to this given pixel, but truly an ubiquitous observation. Again, we stress that this issue is GEO-specific; in comparison, such imperfections are inconsequential for an instrument like CERES as they are simply averaged out, since there is no issue of fixed viewing angles for instruments on LEO orbits.

\begin{figure}[ht]
\centering

	\includegraphics[height=5.7375cm]{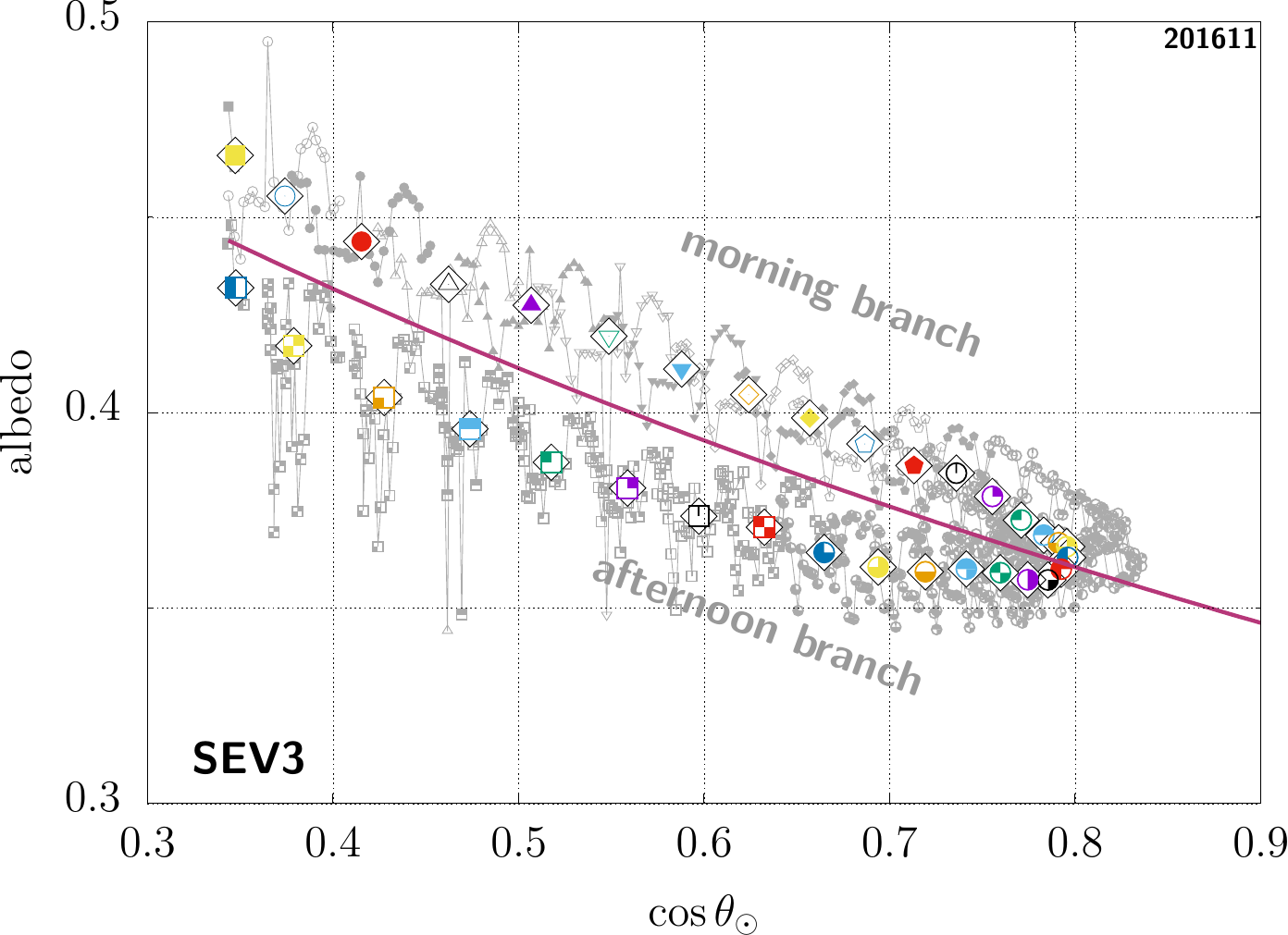}
	\includegraphics[height=5.7375cm, trim = 20 0 0 0, clip]{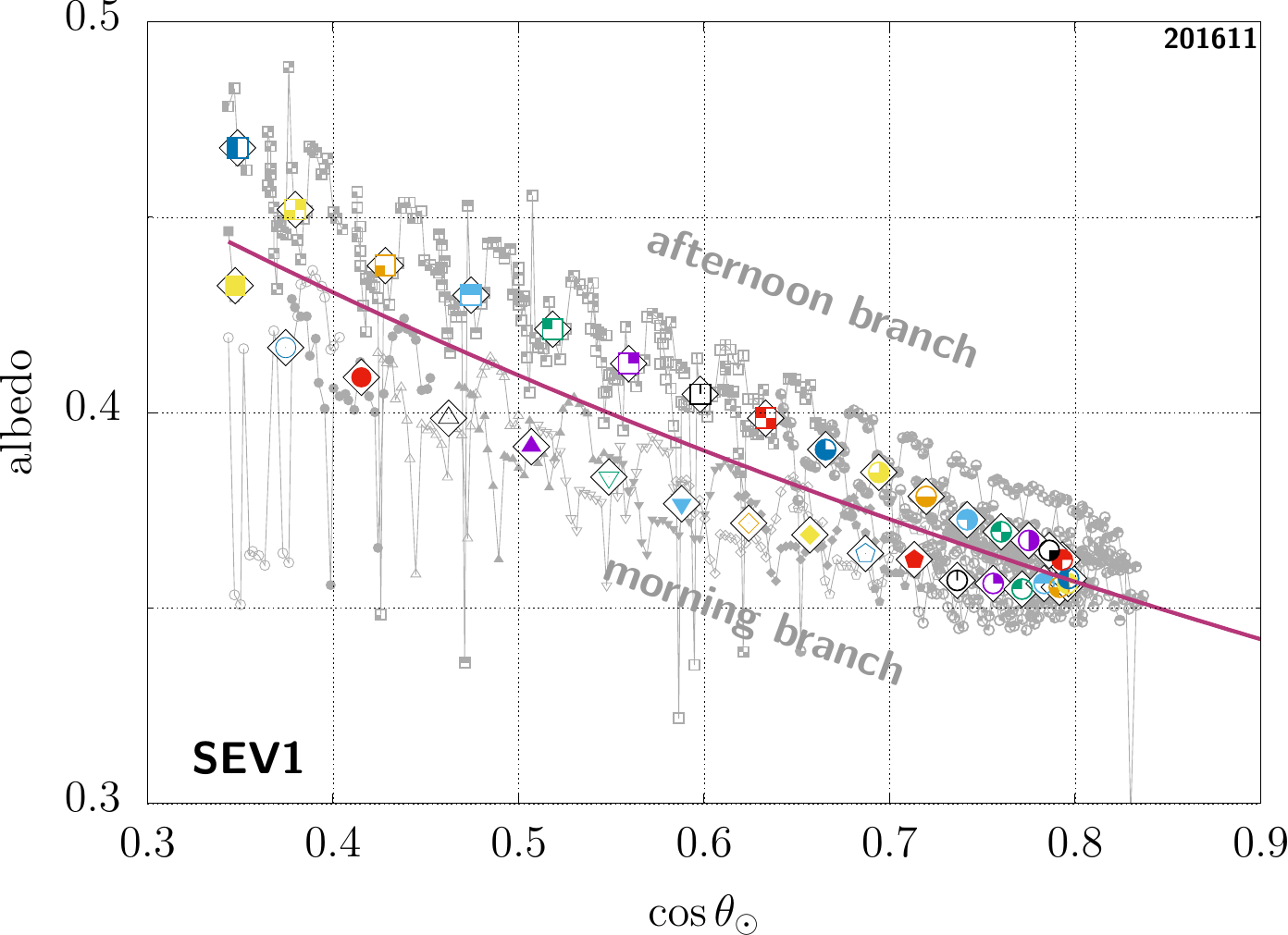}
\caption{Same as Figure~\ref{fig:samplepixel_sahara}, but with the corresponding fit given in Eq.~\eqref{eq:albedo_functional_form} now overlaid in both cases.}
\label{fig:samplepixel_sahara_fit}
\end{figure}

\begin{figure}[ht]
\centering

	\includegraphics[height=5.7375cm]{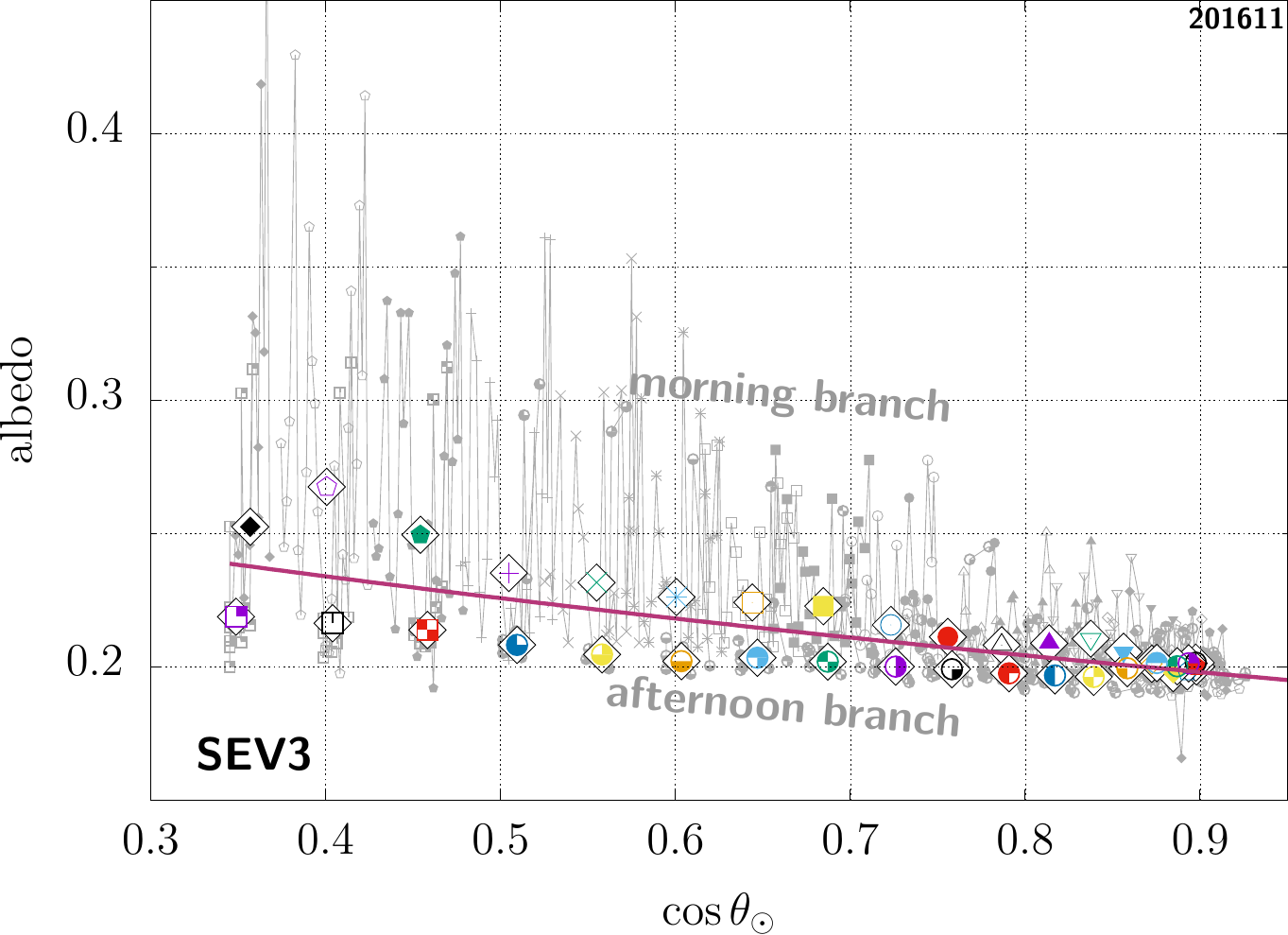}
	\includegraphics[height=5.7375cm, trim = 20 0 0 0, clip]{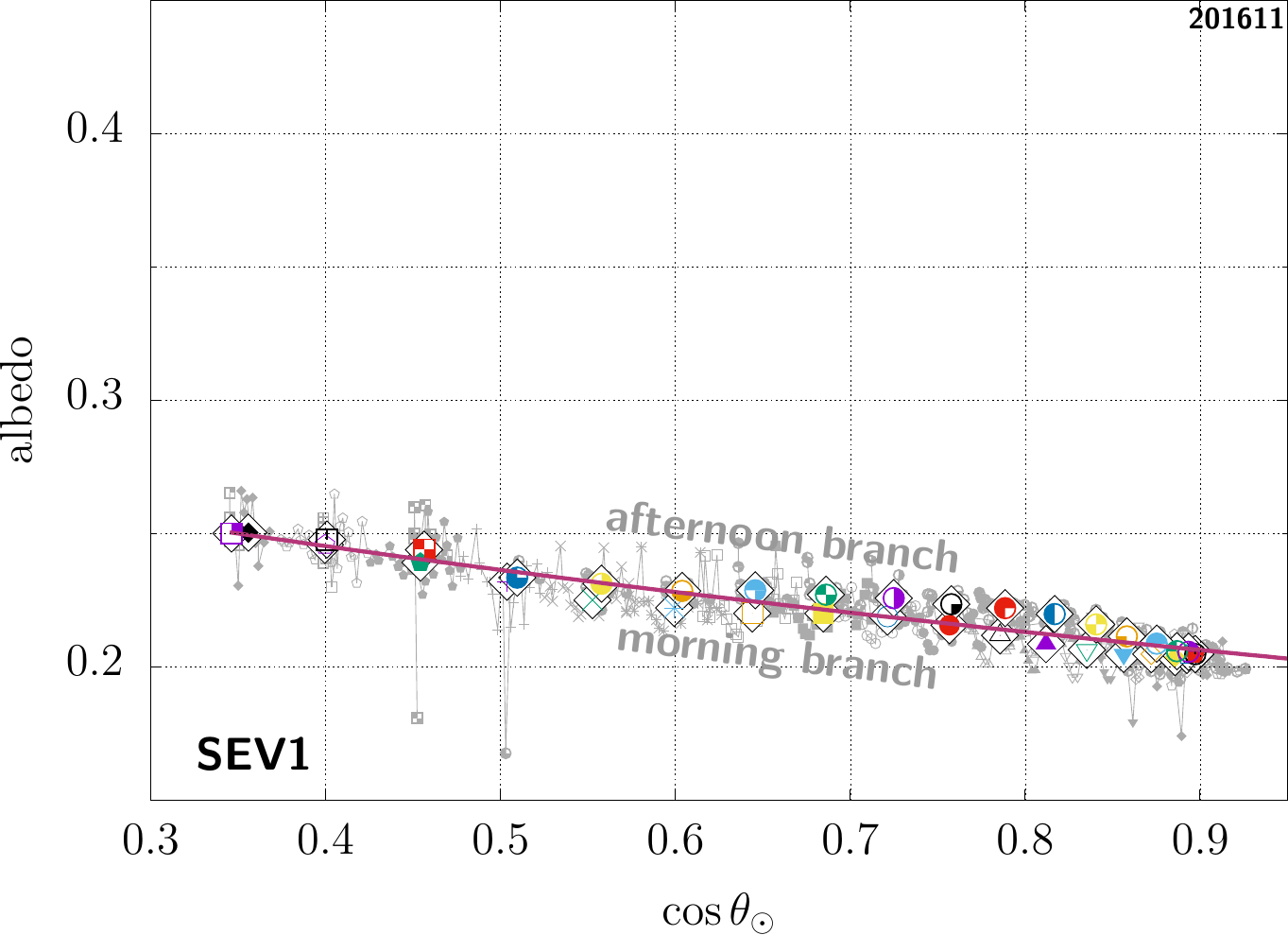}
\caption{Same as Figure~\ref{fig:samplepixel_sahara_fit}, but for another sample 
GEO pixel (latitude 7.45$\degree$, longitude 48.70$\degree$), this time in the Horn of Africa.}
\label{fig:samplepixel_hornofafrica_fit}
\end{figure}

Since the existence of these branches is not physical\footnote{Further note that any actual observer-independent physical effect with a significant impact on the albedo should remain clearly visible even with swapped branches from \MSG{3} to \MSG{1}. If there were dew~\cite{Minnis_etal:1997} in the early morning for instance, the shape of the two morning branches would then be similarly skewed upwards and the apparent symmetry seen here when swapping the morning and afternoon branches would be lost.}, one thing that we can do is to fit the albedo as a function of the solar zenith angle according to the functional form of Eq.~\eqref{eq:albedo_functional_form}, and then use that to calculate the representative mean or median albedo, before projecting. 
These fits are done by non-linear least-squares optimisations taking into account all the available information over the $n_{t_x}$ timeslots:\footnote{This is done using the NLopt library~\cite{NLopt}. Having tested several of the available algorithms, we find that the derivative-free \texttt{NEWUOA}~\cite{LN_NEWUOA} and gradient-based \texttt{SLSQP}~\cite{LD_SLSQP} algorithms are especially reliable, fast, and accurate for the problem at hand.}

\begin{equation}
	\min_{a_{60}, d} \left(  \sum_{t=1}^{n_{t_x}} \left[  a_t - a_{60}\frac{1+d}{1+2d\mu_t}  \right]^2  \right) \qquad \textrm{independently for each GEO pixel};
\end{equation}
where for each timeslot $t$, $\mu_t=\cos\theta_{\odot,t}$ and $a_t$ are respectively the instantaneous values for the cosine of the solar zenith angle and for the albedo derived from the GL-SEV products.
We do stress that the objective here is not to provide a predictive model of what the albedo might be, based on the scene; rather, this is merely an empirical fit, using only all the data for a given pixel over a complete month.
In the case of our example, the results are shown in Figure~\ref{fig:samplepixel_sahara_fit}.
Figure~\ref{fig:samplepixel_hornofafrica_fit} shows another sample pixel, this time in the Horn of Africa and with remaining unidentified clouds from the \MSG{3} viewpoint, especially in the morning.
Note that rather similar results are obtained when using the mean instead of the median albedo, but 
the fit is then necessarily more sensitive to the presence of unidentified clouds, systematically leading to a larger $a_{60}$ and therefore larger albedo values as a function of $\cos(\theta_\odot)$ in those cases.
Another interesting thing to notice is that, as this second example pixel is close to being at the same longitude as the \MSG{1} satellite, seen from there, the branches are then not so spread but close to being indistinguishable from one another.

\begin{figure}[h]
\centering

	\includegraphics[height=5.7375cm, trim = 0 0 0 0, clip]{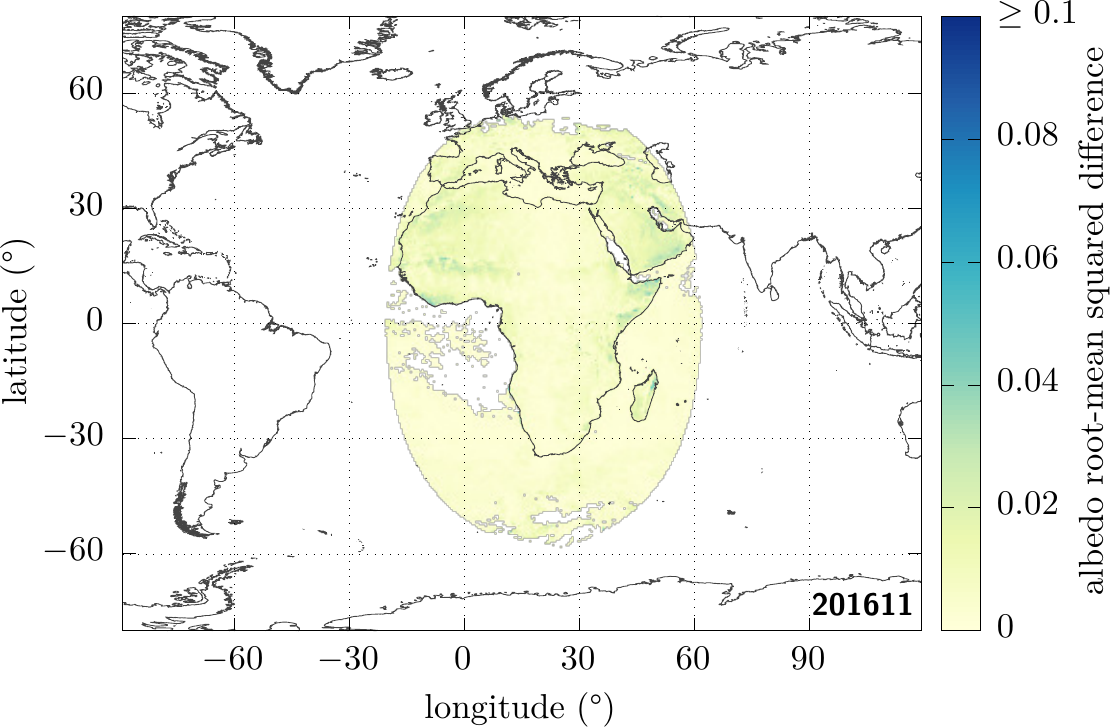}

\caption{Same as Figure~\ref{fig:latlon_orig} (\textbf{d}), but now with imposed angular consistency.}
\label{fig:latlon_corr}
\end{figure}

\begin{figure}[h!]
\centering

	\includegraphics[height=5.205cm, trim = 0 0 15 0, clip]{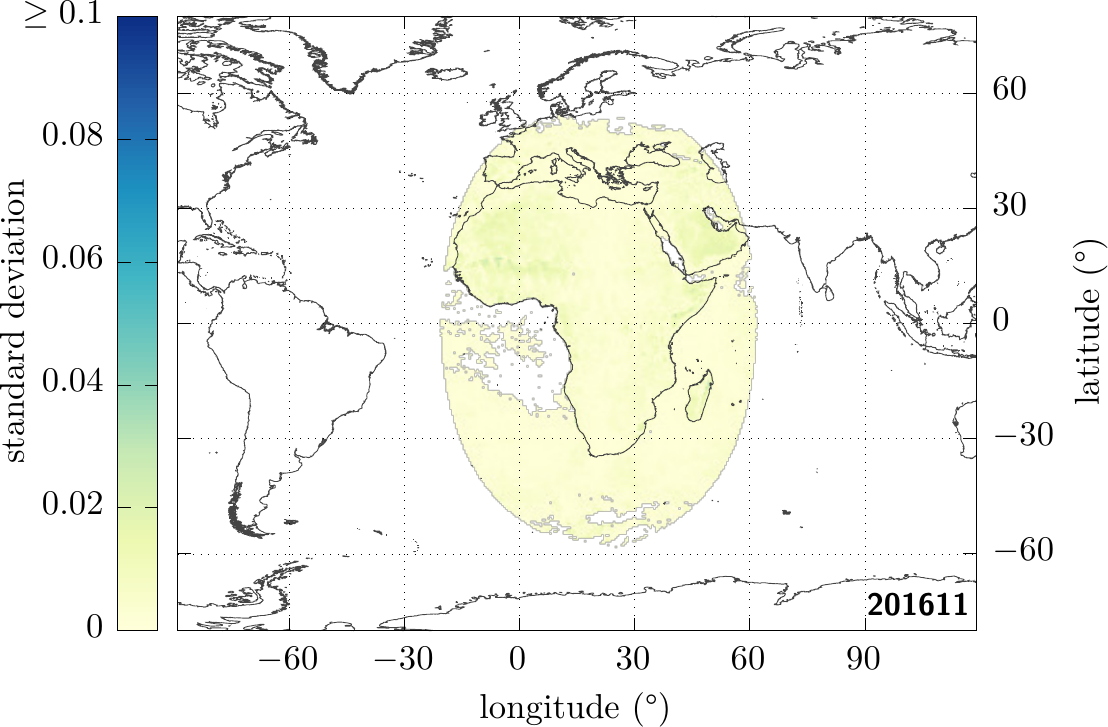}
	\hfill
	\includegraphics[height=5.205cm, trim = 0 0 0 0, clip]{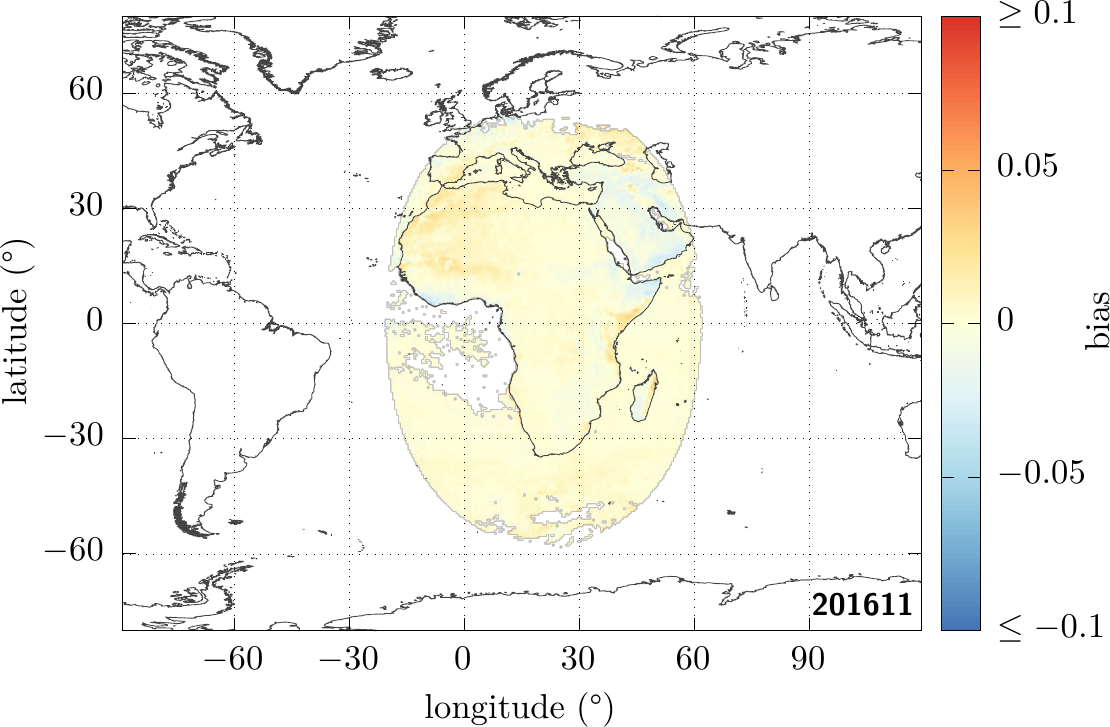}

\caption{Decomposition of Figure~\ref{fig:latlon_corr} in terms of standard deviation (\emph{left}) and bias (\emph{right}).}
\label{fig:latlon_sigmabiascorr}
\end{figure}

In Figure~\ref{fig:latlon_corr}, we show the root-mean square difference in clear-sky albedo retrieved from SEV3 and SEV1, where we both use all the masks that were introduced in Section~\ref{sec:masks} and the imposed angular consistency discussed here 
before projecting onto the latitude--longitude grid.
We request at least three timeslots with defined albedo in order to make the fit and otherwise mask the corresponding pixel.\footnote{One could also further include a minimum requested range in $\cos(\theta_\odot)$, especially for high latitudes in winter.} 
The corresponding summary statistics over all defined pixels are given in Table~\ref{table:201611_withmasksandfits}. In particular, we can see that the quadratic mean of the RMSD over the whole image is now as low as 0.008 in the overlapping region, and that no large discrepancy remains.
Within the RMSD, the standard deviation shown in Figure~\ref{fig:latlon_sigmabiascorr} (\emph{left}) is much decreased as a result of the fit (compare to Figure~\ref{fig:latlon_sigmabiasorig} (\emph{left})). Notice that the bias shown in Figure~\ref{fig:latlon_sigmabiascorr} (\emph{right}) remains small, even though the average absolute bias is slightly larger overall after applying the fit: over the northern part of the African continent and the Arabian peninsula in particular we can see a slightly strengthened trend with a positive bias (light red colour) to the west of $20.75\degree$E (halfway between $0\degree$ and $41.5\degree$E) and a negative bias to the east (light blue colour).

\begin{table}[h!]
\caption{Summary statistics; compare to Table~\ref{table:201611_withmasks}.}
\centering
\begin{tabular}{lccc}
\toprule
&$\bm{\sqrt{\langle\mathrm{RMSD^2}\rangle}}$	&$\bm{\sqrt{\langle\sigma^2\rangle}}$	& $\bm{\langle\mathrm{|bias|}\rangle}$\\
\midrule
with all the masks + empirical fits &	0.008		& 0.004		& 0.005 \\
\bottomrule
\end{tabular}
\label{table:201611_withmasksandfits}
\end{table}

Note that, alternatively, one could instead already proceed with a simple binning in $\cos\theta_\odot$. An approach of this type was actually adopted in a preliminary study with just two bins~\cite{Dewitte_etal:2017}, and the results obtained for the same month ($\sqrt{\langle\mathrm{RMSD^2}\rangle}$ = 0.012, $\sqrt{\langle\sigma^2\rangle} = 0.008$, and $\langle\mathrm{|bias|}\rangle = 0.005$) are comparable to those we report here in Table~\ref{table:201611_withmasksandfits}, giving confidence that the general method itself is quite robust.

\subsection{Merged overhead albedo}

\begin{figure}[ht]
\centering

	\includegraphics[height=5.079375cm, trim = 0 24 54 0, clip]{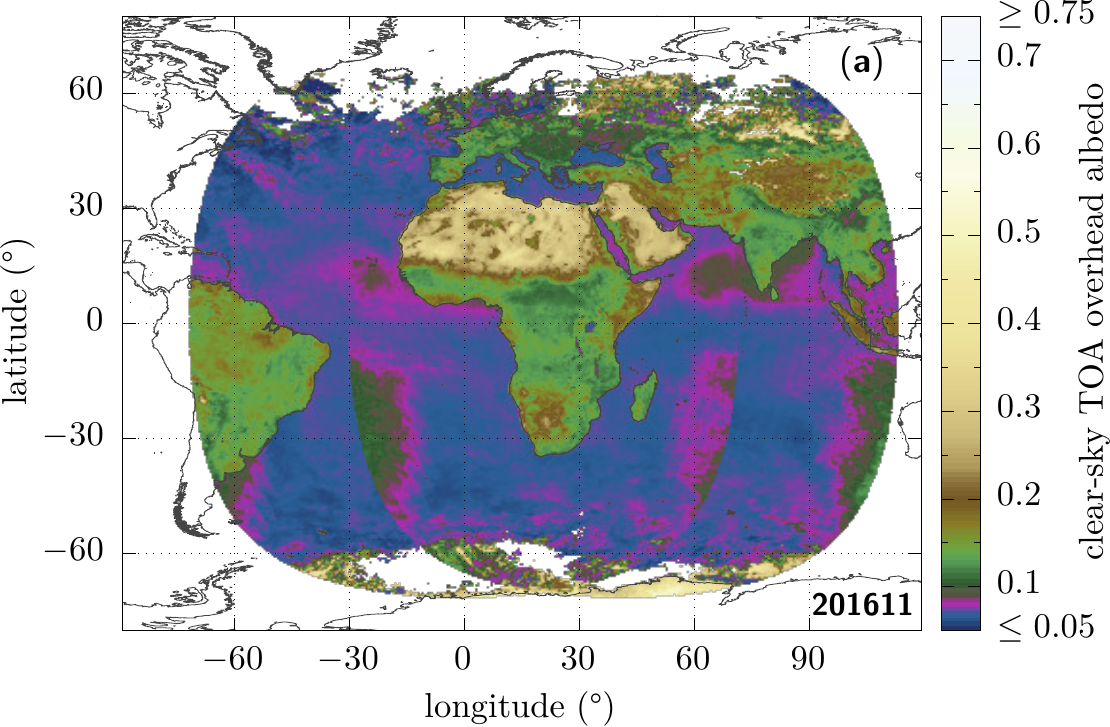} 
	\includegraphics[height=5.079375cm, trim = 35 24 0 0, clip]{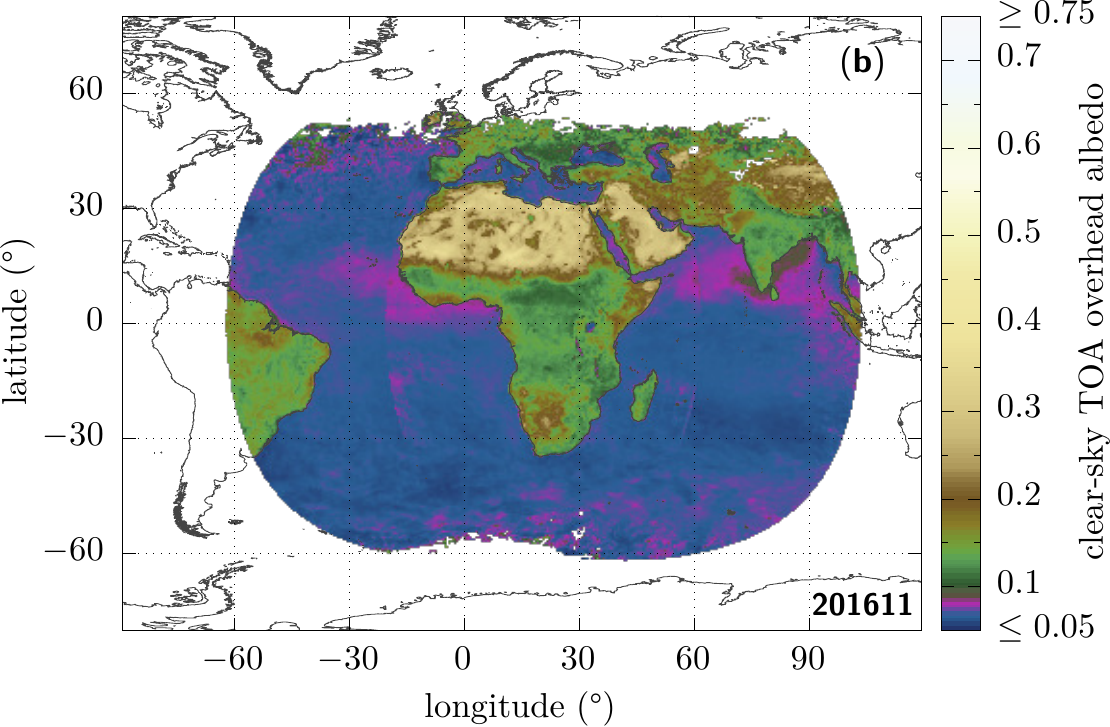} 
	\includegraphics[height=5.7375cm, trim = 0 0 54 0, clip]{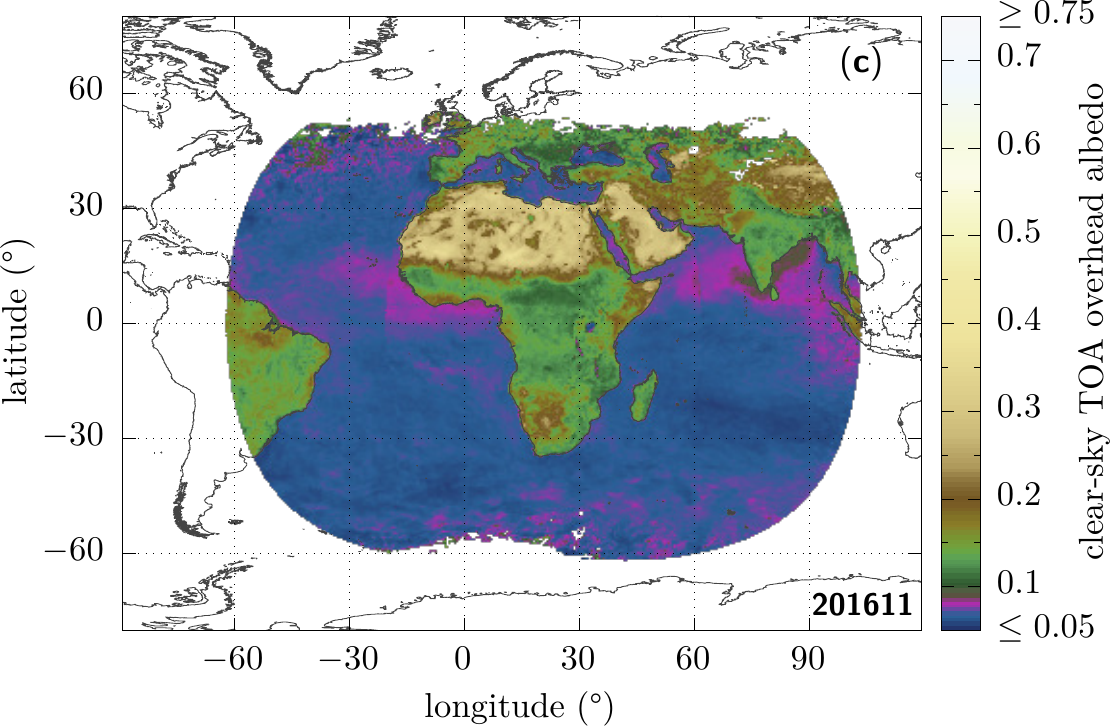} 
	\includegraphics[height=5.7375cm, trim = 35 0 0 0, clip]{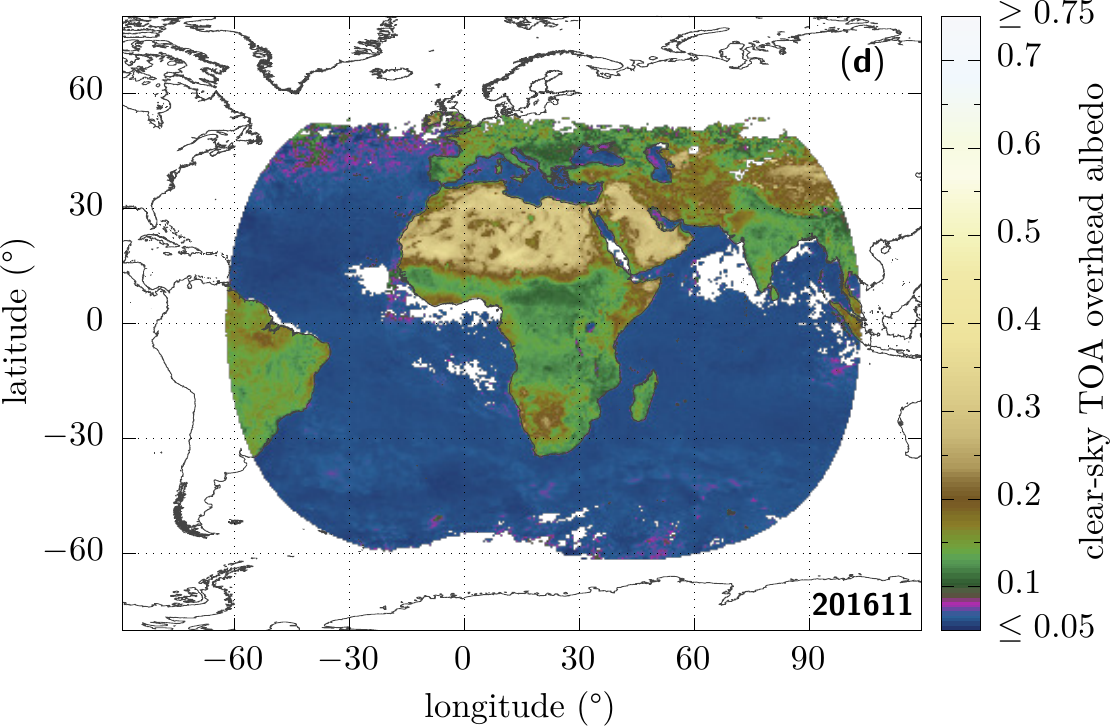}

\caption{Overhead albedo, shown in the different masked cases of Figure~\ref{fig:latlon_orig} after imposing the angular consistency given in Eq.~\eqref{eq:albedo_functional_form} throughout the month: (\textbf{a}, \emph{top left}) no mask; (\textbf{b}, \emph{top right}) zenith-angle masks; (\textbf{c}, \emph{bottom left}) additional sunglint mask; (\textbf{d}, \emph{bottom right}) additional aerosol mask (final result).}
\label{fig:merged_albedooverhead}
\end{figure}

Using false colour rendering, Figure~\ref{fig:merged_albedooverhead} shows the merged overhead albedo $a_0$ in the different cases presented in Figure~\ref{fig:latlon_orig}.
The colours were chosen so that typical values over the ocean ($a_0 \sim 0.06$) would be blue; vegetation ($a_0 \sim 0.12$), green; dark and bright deserts ($a_0 \sim 0.24$ and $\sim 0.36$), brown and yellow. Aerosol-loaded regions over the ocean (with typical $a_0 \sim 0.08$) appear in purple, and can be seen on panel (\textbf{c}) close to India and the Gulf of Guinea in particular.
For each pixel, we used the fitted parameters obtained when imposing the functional form discussed in the previous section. For an overhead Sun, $a_0$ can be simply calculated from

\begin{equation}
	a_0 = a(\theta_\odot = 0) = a_{60} \frac{1+d}{1+2d}.
	\label{eq:albedooverhead_functional_form}
\end{equation}
Whenever the albedo is defined for both SEV3 and SEV1 for a given latitude--longitude pixel, an average is taken.\footnote{A few aerosol-loaded pixels remain west of Africa in the dual-view region, even in Figure~\ref{fig:merged_albedooverhead} (\textbf{d}).
Correctly masked when seen from \MSG{3}, they were not identified from \MSG{1} (being close to the viewing-zenith limit), and thus reintroduced.}
The original sharp discontinuities between satellite views are significantly decreased as the masks described in Sec.~\ref{sec:masks} (zenith-angles (\textbf{b}), sunglint (\textbf{c}), and aerosols (\textbf{d})) are being applied.

\subsection{Visualising the diurnal-asymmetry artefact from the viewpoint of each satellite}\label{sec:visualising_artefact}

\begin{figure}[h]
\centering
	\includegraphics[height=5.7375cm]{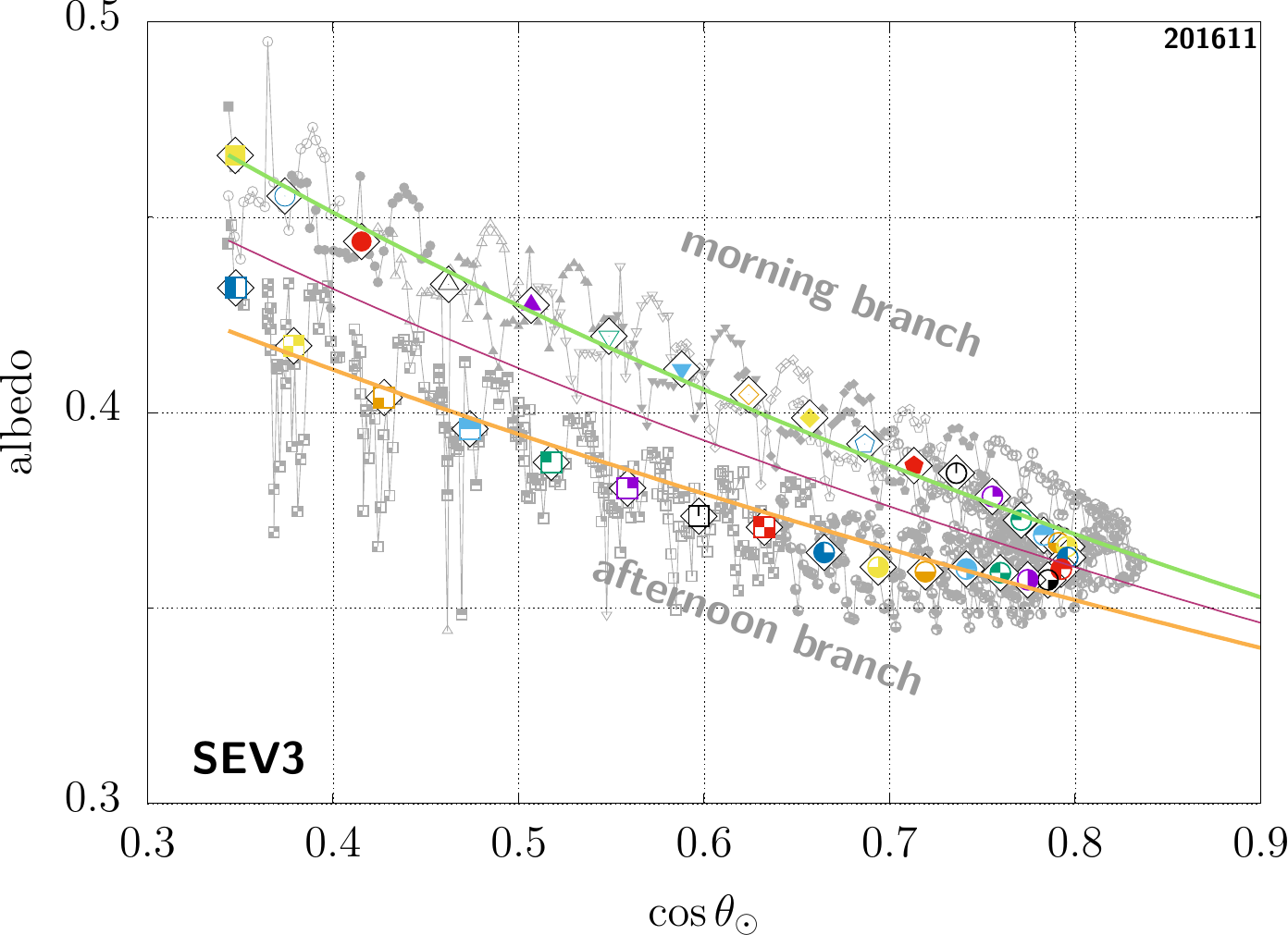}
	\includegraphics[height=5.7375cm, trim = 20 0 0 0, clip]{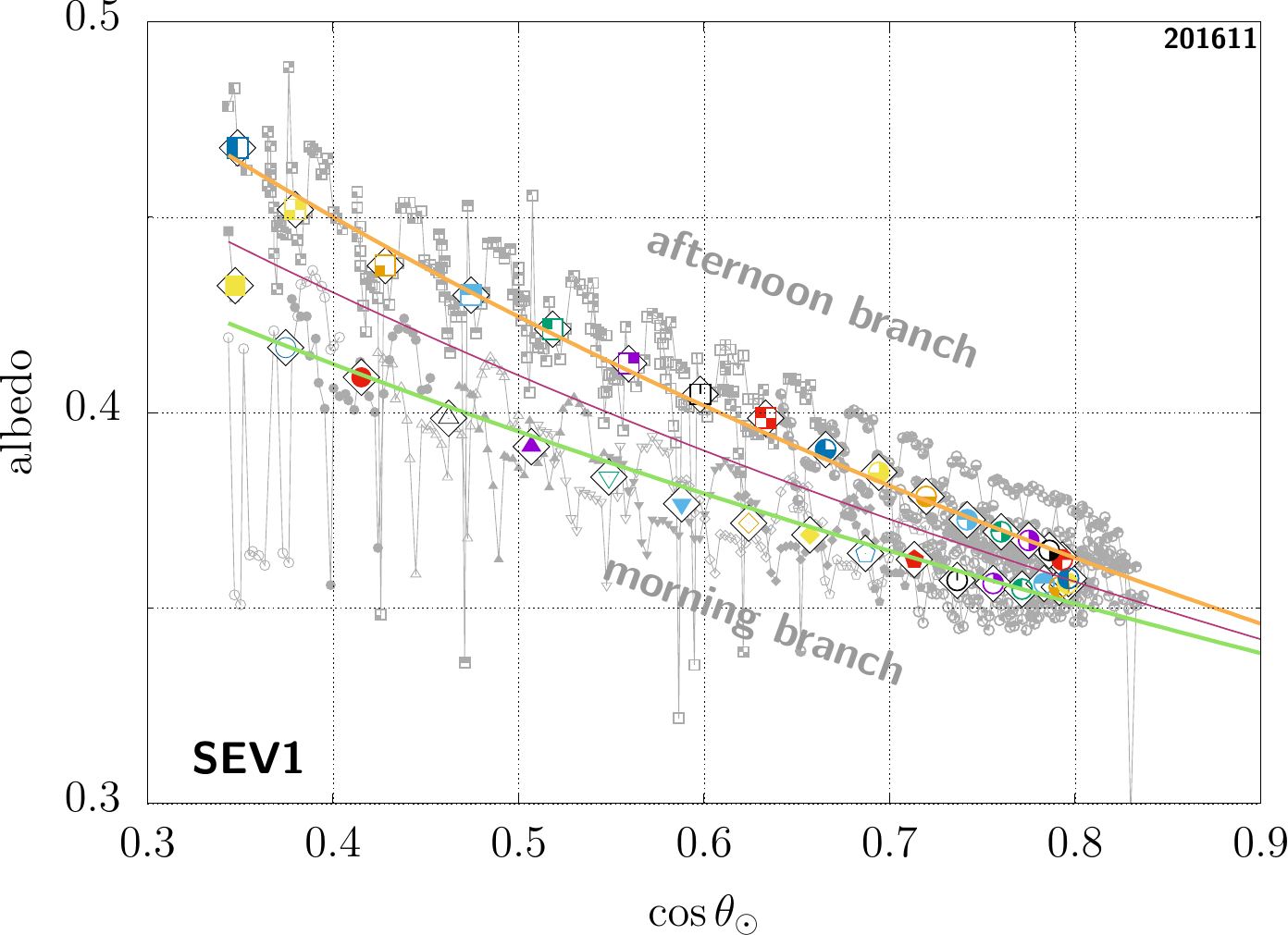}
\caption{Same case as Figure~\ref{fig:samplepixel_sahara_fit}, with each branch fitted 
to enable
the visualisation shown in Figure~\ref{fig:latlon_morningafternoonbranches}.}
\label{fig:illustr_branches_fits}
\end{figure}

Before applying the method described above to a full year of data, let us take a closer look at the spatial distribution of the observer-dependent diurnal-asymmetry artefact. 
It is indeed interesting to be able to visualise it, not only one pixel at a time, but for all pixels at once, from the viewpoint of each satellite.
We are going to fit the morning and afternoon branches separately and compare them.\footnote{For each pixel, we define the branches themselves with respect to the local noon, determined as the UTC timeslot with the largest mean $\cos(\theta_\odot)$ value (akin to Ref.~\cite{Bertrand_etal:2006_asym}); the local midnight is calculated as local noon plus 12 hours, modulo 24 hours.} This will highlight where, according to each satellite, morning albedos appear larger than afternoon albedos and \textit{vice versa}.
Though the shapes of the individual branches are not strictly given by Eq.~\eqref{eq:albedo_functional_form}, this fitting procedure will provide instructive results.
Figure~\ref{fig:illustr_branches_fits} shows at pixel-level the result of such fits for both SEV3 and SEV1, for the rarely cloudy Sahara location discussed in Sec.~\ref{sec:diurnal_asymmetry}. 
To keep as many pixels as possible while making sure that each branch is well-sampled and not overly sensitive to the curvature change close to the tip around the local noon, we find that 
a good trade-off is to only keep pixels with non-masked data spanning at least 12 different timeslots for each branch.\footnote{Due to the tip, we can expect higher branches to be slightly steeper, and lower branches, flatter. While this can artificially increase the difference between branches for each pixel, it will not affect the qualitative result (\textit{i.e.\@} which is higher or lower).}

\begin{figure}[h!]
\centering
	\includegraphics[height=5.7375cm, trim = 0 0 64 0, clip]{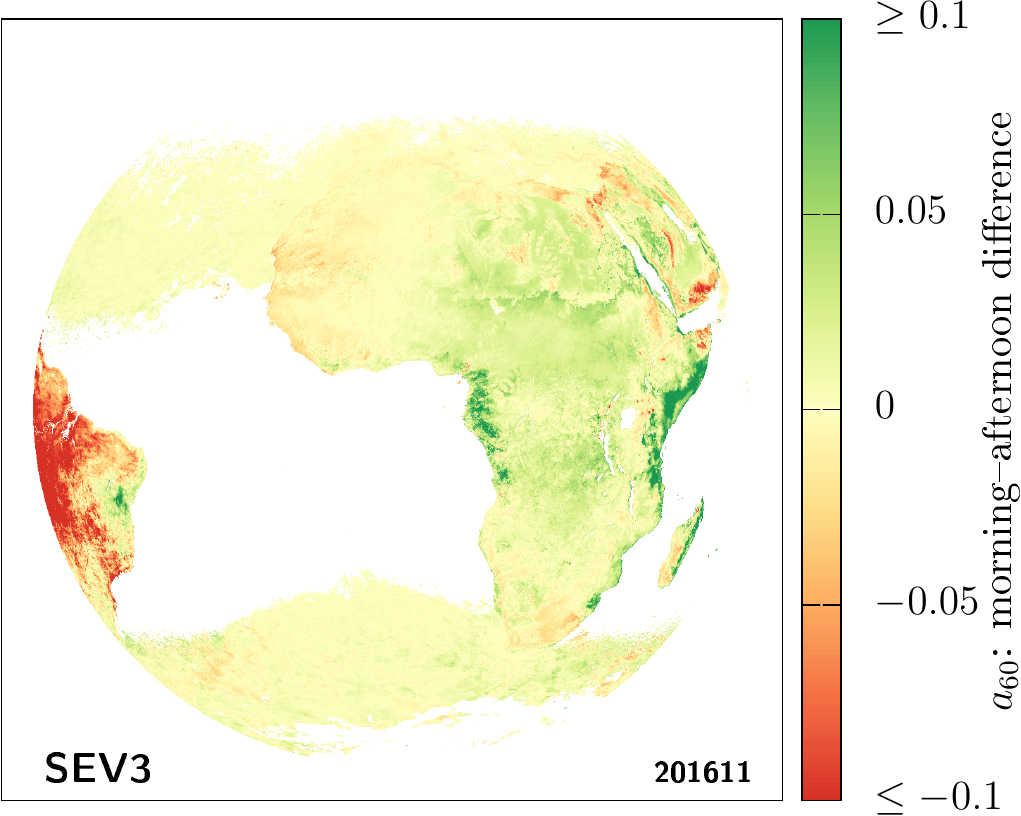}
	\includegraphics[height=5.7375cm, trim = 0 0 0 0, clip]{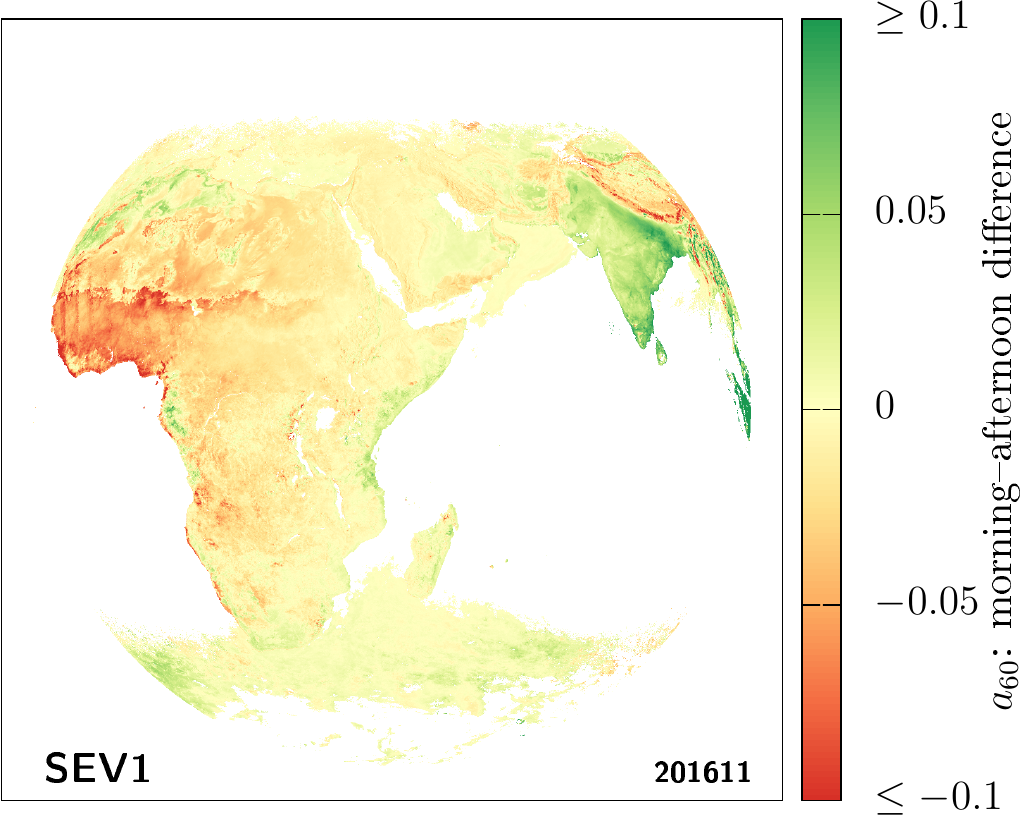}
\caption{Visualisation of the diurnal asymmetry artefact. This compares for each pixel the $a_{60}$ fit parameters (albedo corresponding to $\theta_\odot = 60\degree$) obtained for each of the branches $\left(a_{60,{\mathrm{AM}}} - a_{60,{\mathrm{PM}}}\right)$. The asymmetry, seen from \MSG{3} (\emph{left}) or \MSG{1} (\emph{right}), is obviously strongly observer dependent.}
\label{fig:latlon_morningafternoonbranches}
\end{figure}

\begin{figure}[h!]
\centering
	\includegraphics[height=5.7375cm, trim = 0 0 50 0, clip]{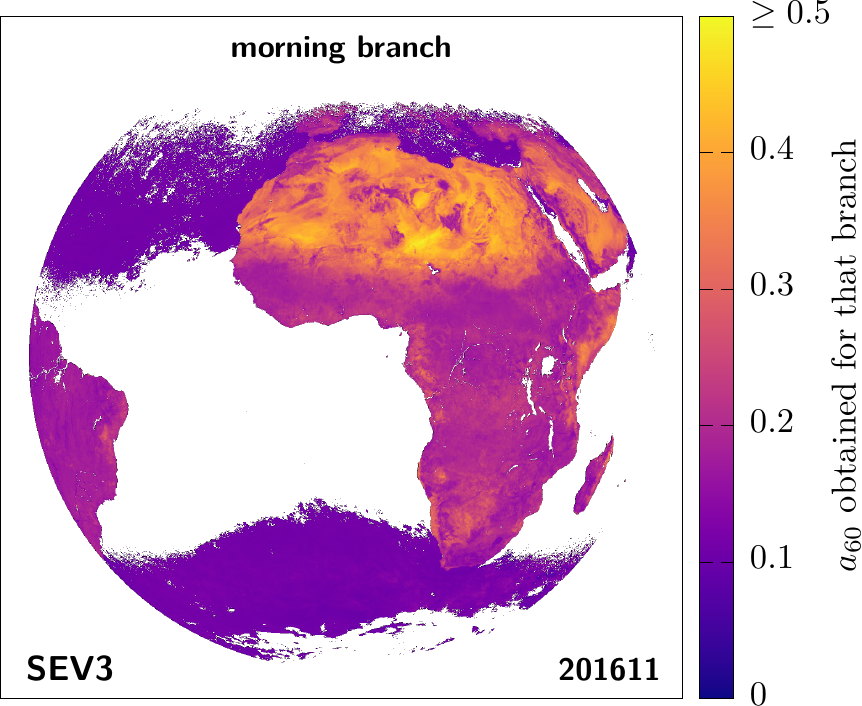}
	\includegraphics[height=5.7375cm, trim = 0 0 0 0, clip]{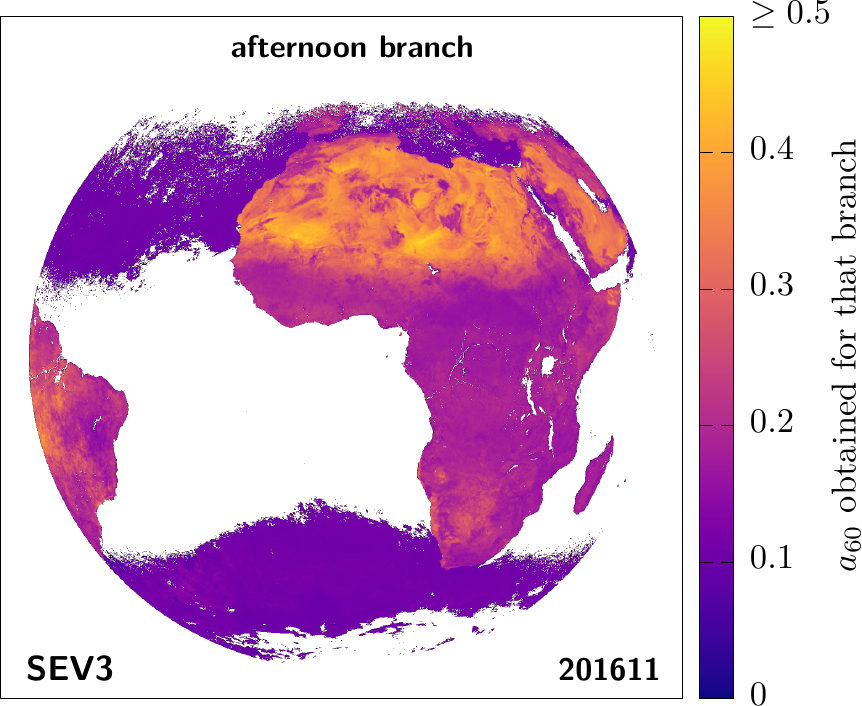}
	
	\includegraphics[height=5.7375cm, trim = 0 0 50 0, clip]{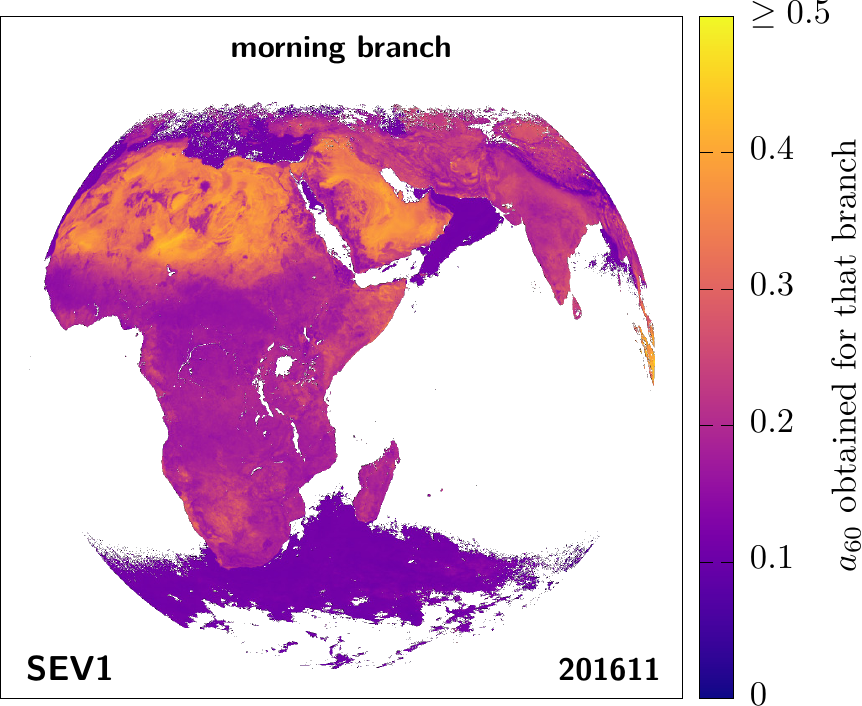}
	\includegraphics[height=5.7375cm, trim = 0 0 0 0, clip]{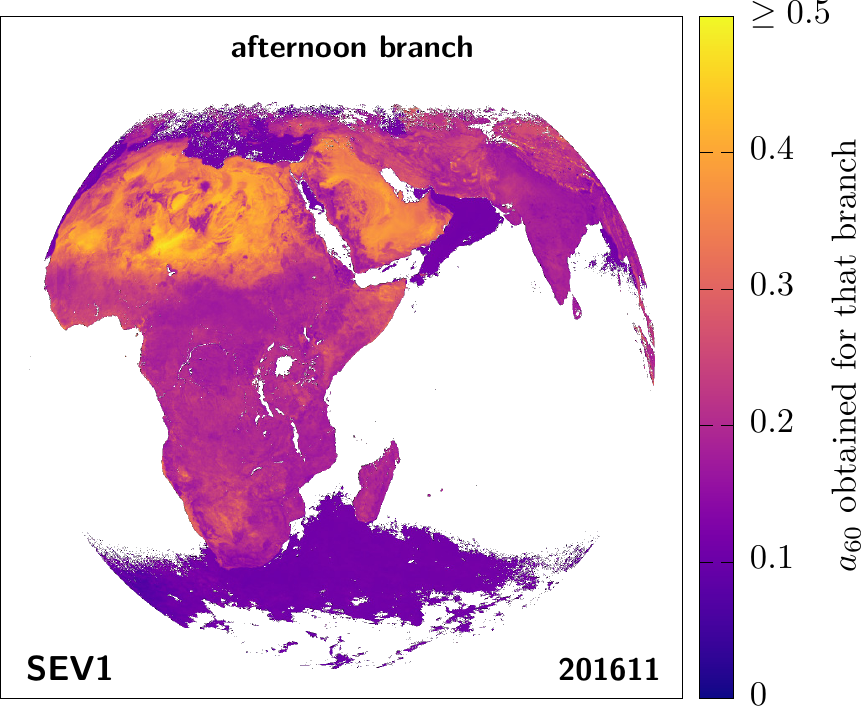}
\caption{Morning and afternoon $a_{60}$-fit-parameter mapping, for both \MSG{3} and \MSG{1}.}
\label{fig:latlon_morningafternoon_a60}
\end{figure}

Figure~\ref{fig:latlon_morningafternoonbranches} then compares for each pixel the $a_{60}$ parameter obtained when fitting the morning branch (AM) to the one obtained for the afternoon branch (PM). The gaps in ocean pixels are largely due to the sunglint and aerosol masks on the ocean, which prevent meeting our fit-quality requirements. 
As we focus on the common region seen from both satellites and centred on Africa, we see that the two satellites strongly disagree on which branch would be larger than the other virtually everywhere. 
From the point of view of \MSG{3} (\emph{left} panel), compared to the satellite position (centre of the image), the albedo tends to be systematically larger during the morning than in the afternoon for pixels on the right of the image (\textit{i.e\@} greener on the right), while the situation tends to be reversed to the left of the image, where the afternoon albedo is then larger than the morning albedo (\textit{i.e\@} redder on the left). In other words, the retrieved albedos for pixels to the east of the satellite tend to be brighter in the morning, and those to the west, brighter in the afternoon, as seen also in Figure~\ref{fig:latlon_morningafternoon_a60}.
Seen from \MSG{1} (\emph{right} panel), there seems to be a similar left/right trend with respect to the satellite position at the centre of the image.
This again highlights that this diurnal asymmetry is a viewing-angle issue.
Note that the rare exceptions where both satellites agree appear to be mostly linked to overimposed cloud-identification issues, such as in Gabon or near the Horn of Africa in the morning (see also Figure~\ref{fig:latlon_morningafternoon_a60}).\footnote{The large afternoon effect in South America is similarly due to a cloud-detection issue.
Note that trying to address it with masks and cuts ultimately creates an imbalance between the branches (artificially giving more weight to one of them). This issue should preferably be dealt with within the GL-SEV data processing itself.}

Without correcting this artefact, SEV3 would tell that most of Africa has a larger albedo in the morning, while SEV1 would tell that it is larger in the afternoon instead.\footnote{As a corollary, this of course also means that, together with Ref.~\cite{Yang_etal:2008}, we do not find that the morning albedo is systematically larger than the afternoon one when such a diurnal asymmetry is present\hspace{1pt}---\hspace{1pt}\textit{i.e.\@} what one might have expected if an actual phenomenon such as dew was at play~\cite{Minnis_etal:1997}.}

\section{One year}
\label{sec:one_year}

Finally, we now simply apply the treatment discussed in Secs.~\ref{sec:masks} and~\ref{sec:diurnal_asymmetry} to each of the months of the year 2017. Again, this is done independently for \MSG{3} and \MSG{1}. 
As before, we compare the results and compute the summary statistics; these are given in Table~\ref{table:one_year}. As expected, the results are comparable to those given in Tables~\ref{table:201611_withmasks} and~\ref{table:201611_withmasksandfits} and the consistency is systematically much improved.
For completeness, note that in the 2017 data which was available for us to compute those, there were six entire days with missing GL-SEV data from \MSG{1}: from January 17th to January 22th (\MSG{1} SEVIRI decontamination), and on November 12th.

\begin{table}[H]
\caption{Summary statistics; compare to Tables~\ref{table:201611_withmasks} and~\ref{table:201611_withmasksandfits}.}
\label{table:one_year}
\centering

\begin{tabular}{cc}

\begin{tabular}{cccc}
&&\makebox[1.5cm]{\textbf{\emph{masks only}}}\\
\toprule
\textbf{month} &$\bm{\sqrt{\langle\mathrm{RMSD^2}\rangle}}$	&$\bm{\sqrt{\langle\sigma^2\rangle}}$	& $\bm{\langle\mathrm{|bias|}\rangle}$\\
\midrule
2017/01&	0.016			& 0.013			& 0.005\\
2017/02&	0.019			& 0.017			& 0.005\\
2017/03&	0.017			& 0.015			& 0.004\\
2017/04&	0.016			& 0.015			& 0.004\\
2017/05&	0.015			& 0.014			& 0.003\\
2017/06&	0.016			& 0.015			& 0.004\\
2017/07&	0.018			& 0.016			& 0.005\\
2017/08&	0.021			& 0.018			& 0.006\\
2017/09&	0.020			& 0.018			& 0.005\\
2017/10&	0.017			& 0.016			& 0.004\\
2017/11&	0.014			& 0.013			& 0.004\\
2017/12&	0.015			& 0.013			& 0.005\\
\bottomrule
\end{tabular}

&

\begin{tabular}{cccc}
&&\makebox[1.5cm]{\textbf{\emph{masks + empirical fits}}}\\
\toprule
\textbf{month} &$\bm{\sqrt{\langle\mathrm{RMSD^2}\rangle}}$	&$\bm{\sqrt{\langle\sigma^2\rangle}}$	& $\bm{\langle\mathrm{|bias|}\rangle}$\\
\midrule
2017/01&	0.011			& 0.006			& 0.006\\
2017/02&	0.011			& 0.005			& 0.005\\
2017/03&	0.010			& 0.006			& 0.004\\
2017/04&	0.009			& 0.006			& 0.004\\
2017/05&	0.009			& 0.006			& 0.004\\
2017/06&	0.009			& 0.006			& 0.005\\
2017/07&	0.009			& 0.005			& 0.005\\
2017/08&	0.011			& 0.006			& 0.006\\
2017/09&	0.011			& 0.006			& 0.006\\
2017/10&	0.009			& 0.005			& 0.005\\
2017/11&	0.009			& 0.005			& 0.004\\
2017/12&	0.010			& 0.005			& 0.005\\
\bottomrule
\end{tabular}

\end{tabular}
\end{table}

Further averaging these monthly results for the quadratic mean of the RMSD, the quadratic mean of the standard deviation, and the mean absolute bias over all the months of 2017 in the masks-only case respectively gives 0.017, 0.015, and 0.005. When empirical fits are also used, one then respectively obtains 0.010, 0.006, and 0.005.

\section{Conclusions}
\label{sec:conclusions}

In this paper, we have had a closer look at the clear-sky top-of-atmosphere broadband shortwave albedo retrieved from MSG geostationary satellites positioned at 0$\degree$ and $41.5\degree$E longitude, and compared the results obtained over the very substantial overlapping region observed by both.

This was first done for the month of November 2016.
We could identify discrepancies which exist at present in the GL-SEV products.
Among the main sources of discrepancies, some in particular were identified to be due to the fluxes retrieved in the sunglint regions over ocean pixels close to the limbs, where the RMSD can locally well exceed values as large as $0.10$; another important source of issues being related to aerosol-loaded regions, where the RMSD can be as large as $\sim 0.05$.
For the aerosol, the observer-dependent differences can be tied to the lack of a dedicated ADM processing.
One of the key points stressed in this study is therefore that the use of ADMs appropriate for the case of `aerosol over ocean' direly needs to be included in the GERB/GL-SEV processing.
We then aimed at improving the situation.
This was achieved first by using masks to remove pixels for which it appears at present difficult to obtain a sufficiently reliable flux, and then by using an empirical physical fit of the expected general dependence of the albedo with the solar zenith angle to apply level-3 corrections at the level of one month\hspace{1pt}---\hspace{1pt}taking advantage of the high temporal resolution characteristic of geostationary satellites.
The imposed angular consistency helped address what clearly turned out to be an unphysical observer-dependent diurnal asymmetry artefact, which we also quite extensively discussed.
Note that at no point was the additional information coming from the dual view used to correct the results from either of the satellites. 
They both remained treated independently throughout.
These led to sizeable improvements in terms of consistency.

The same approach was then applied over one full year, for all the months of 2017, and we similarly obtained that the agreement between the retrieved albedos from each satellite at the level of one month is much improved with this method.
In average, over the twelve months of 2017, the quadratic mean of the albedo root-mean squared difference indeed goes down to 0.01, while the corresponding average quadratic mean of the standard deviation and average mean absolute bias are 0.006 and 0.005 respectively.

The determination of the radiative fluxes from GEO-satellite instruments is needed to resolve the diurnal cycle of the ERB. The results of our paper could be used to improve global ERB products, based on a combination of LEO measurements with a ring of GEO instruments. With such a GEO ring, there could be multiple regions where overlapping GEO instruments provide a dual-view validation of the estimated radiative fluxes.


\vspace{6pt} 



\authorcontributions{conceptualisation, S.D. and A.P.; methodology, A.P.; software, A.P.; formal analysis, A.P.; validation, S.D. and N.C.; investigation, A.P., S.D. and N.C.; writing---original draft preparation, A.P.; writing---review and editing, N.C. and S.D; visualisation, A.P.
All authors have read and agreed to the published version of the manuscript.}

\funding{This research was funded by the Solar-Terrestrial Centre of Excellence (STCE).  The STCE is a collaboration between the Royal Meteorological Institute of Belgium (RMIB), the Royal Observatory of Belgium (ROB), and the Belgian Institute for Space Aeronomy (BISA).}

\acknowledgments{
It is our pleasure to thank Christine Aebi for a careful reading of and various comments on the manuscript.
We also thank the four anonymous reviewers, who contributed to improve our paper through their useful critical and dedicated feedback.
}

\conflictsofinterest{The authors declare no conflict of interest. The funders had no role in the design of the study; in the collection, analyses, or interpretation of data; in the writing of the manuscript, or in the decision to publish the results.
} 

\abbreviations{The following abbreviations are used in this manuscript:\\

\noindent 
\begin{tabular}{@{}ll}
ADM & Angular Distribution Model\\
CERES & Clouds and the Earth's Radiant Energy System\\
CM SAF & Satellite Application Facility on Climate Monitoring\\
ERB & Earth's Radiation Budget\\
ESA & European Space Agency\\
EUMETSAT & European Organisation for the Exploitation of Meteorological Satellites\\
GCOS & Global Climate Observing System\\
GEO & Geostationary Orbit\\
GERB & Geostationary Earth Radiation Budget\\
GL-SEV & SEVIRI `GERB-like' synthetic product\\
HDF & Hierarchical Data Format\\
LEO & Low Earth Orbit\\
MMDC & Monthly Mean Diurnal Cycle\\
MSG & Meteosat Second Generation\\
NASA & National Aeronautics and Space Administration\\
RMSD & (Albedo) Root-Mean Squared Difference\\
SEVIRI & Spinning Enhanced Visible and InfraRed Imager\\
SYN1deg & Synoptic 1$\degree$\\
TOA & Top of Atmosphere\\
TRMM & Tropical Rainfall Measuring Mission\\
UTC & Coordinated Universal Time
\end{tabular}}

\appendixtitles{no} 
\appendix
\section{} 
\label{app:patterns}

\begin{figure}[h!!!]
\centering

	\includegraphics[width=.85\textwidth]{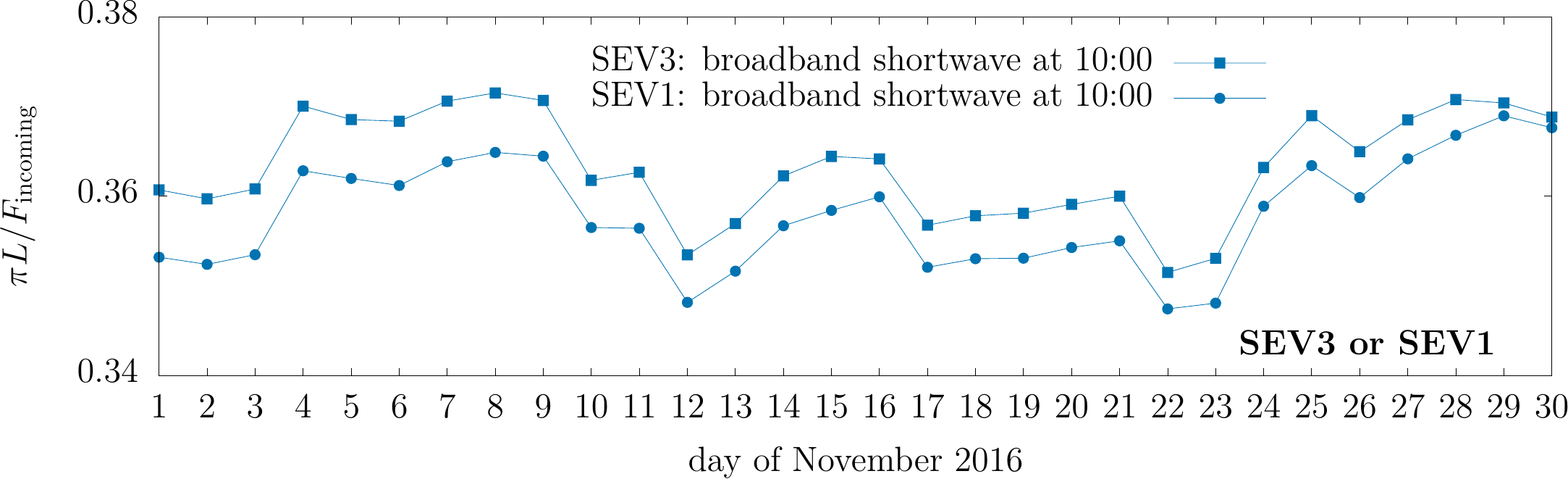}
	
\caption{Consistent patterns in instantaneous data seen from \MSG{3} (squares) and \MSG{1} (dots). This shows the pseudo-albedo $\pi L/F_{\rm incoming}$, calculated directly from the broadband GL-SEV radiances at 10:00 UTC, for every day of the month of 2016/11 in the Sahara sample case shown in Figure~\ref{fig:samplepixel_sahara}. Note that in this case the pattern is inverted along the $x$-axis when shown as a function of the day of the month instead of $\cos\theta_\odot$ (\textit{i.e.\@} $\cos\theta_\odot$ values are larger at the beginning of the month, smaller at the end).}
\label{fig:patterns_GL}
\end{figure}

Here, we show that the patterns visible in Figure~\ref{fig:samplepixel_sahara} in the instantaneous albedo data at each timeslot are already present in the broadband GL-SEV radiances. 
To remove any influence from the ADMs, Figure~\ref{fig:patterns_GL} shows this time the pseudo-albedo as a function of the day of the month for the 10:00 UTC timeslot.
Since this is independent of the ADMs, there is no angular correction applied at all for the surface type with respect to an ideal Lambertian surface; it is therefore expected that the exact pseudo-albedo values from \MSG{3} or \MSG{1} cannot be directly meaningfully compared.
What matters however is that a similar pattern/shape is clearly seen from each satellite over that timeslot. 

\begin{figure}[h!!!]
\centering

	\includegraphics[width=.85\textwidth]{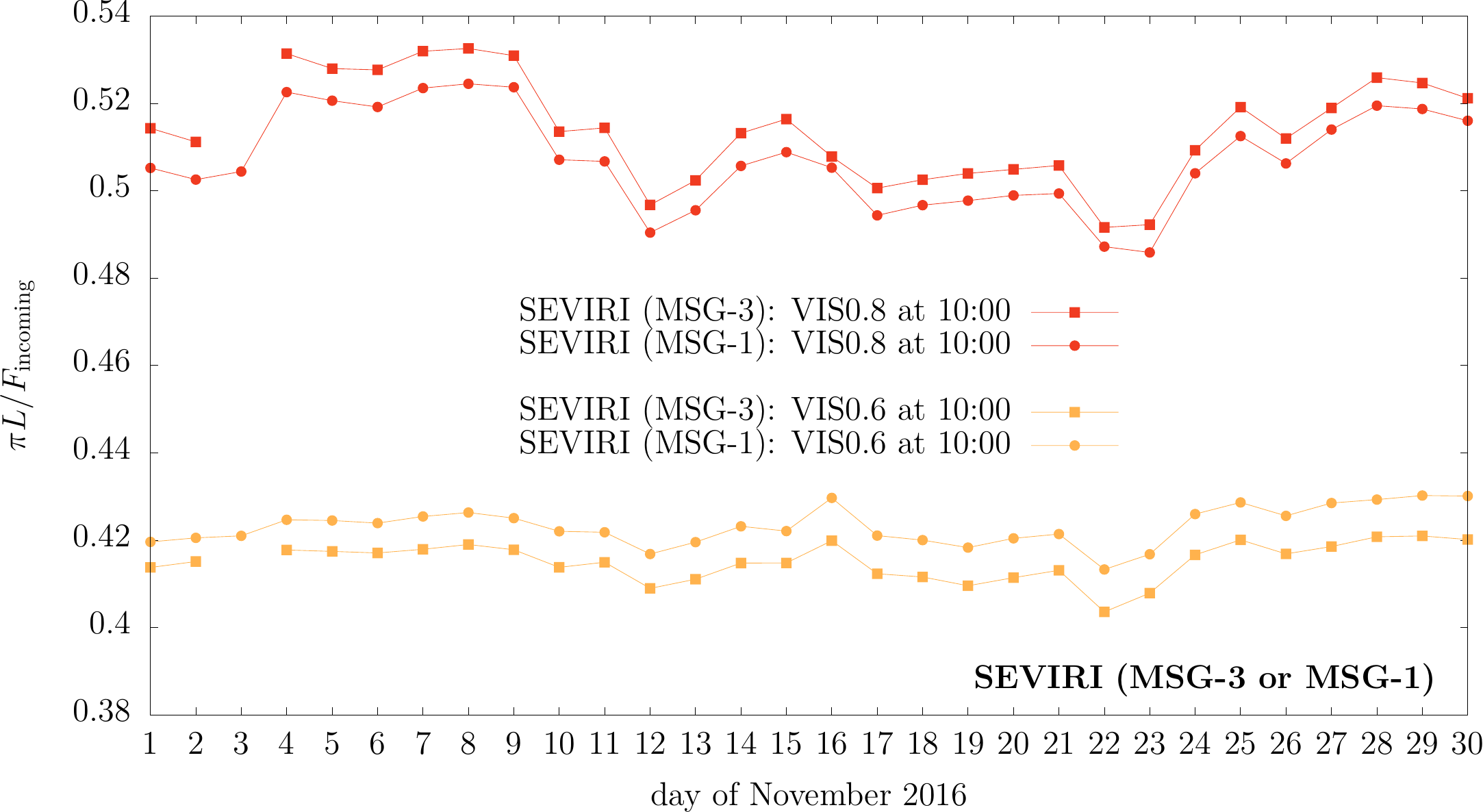}
	
\caption{Same as Figure~\ref{fig:patterns_GL}, but this time using directly SEVIRI narrowband radiance data for the corresponding pixels at 10:00 UTC, in the same $3\times3$ (S3) grid as the GL-SEV products.}
\label{fig:patterns_SEVIRI_NB}
\end{figure}

Such a pattern is actually even visible in the SEVIRI narrowband radiance data and the two independent satellites again appear to agree; these fluctuations are therefore very likely physical. Figure~\ref{fig:patterns_SEVIRI_NB} shows the VIS0.8 and VIS0.6 channels, and VIS0.8 in particular is quite clearly reminiscent of the pattern in broadband GL-SEV synthetic data.

Let us stress that the presence of such patterns is a general observation, not limited to this specific example case (though the pattern shape itself of course changes for different pixels and periods of time).
They can sometimes surprisingly remain essentially unchanged at different times of the day, as is the case for the current Sahara example (seemingly same pattern repeated over different timeslots).

\section{} 
\label{app:gerb}

\begin{figure}[ht]
\centering

	\includegraphics[height=5.7375cm]{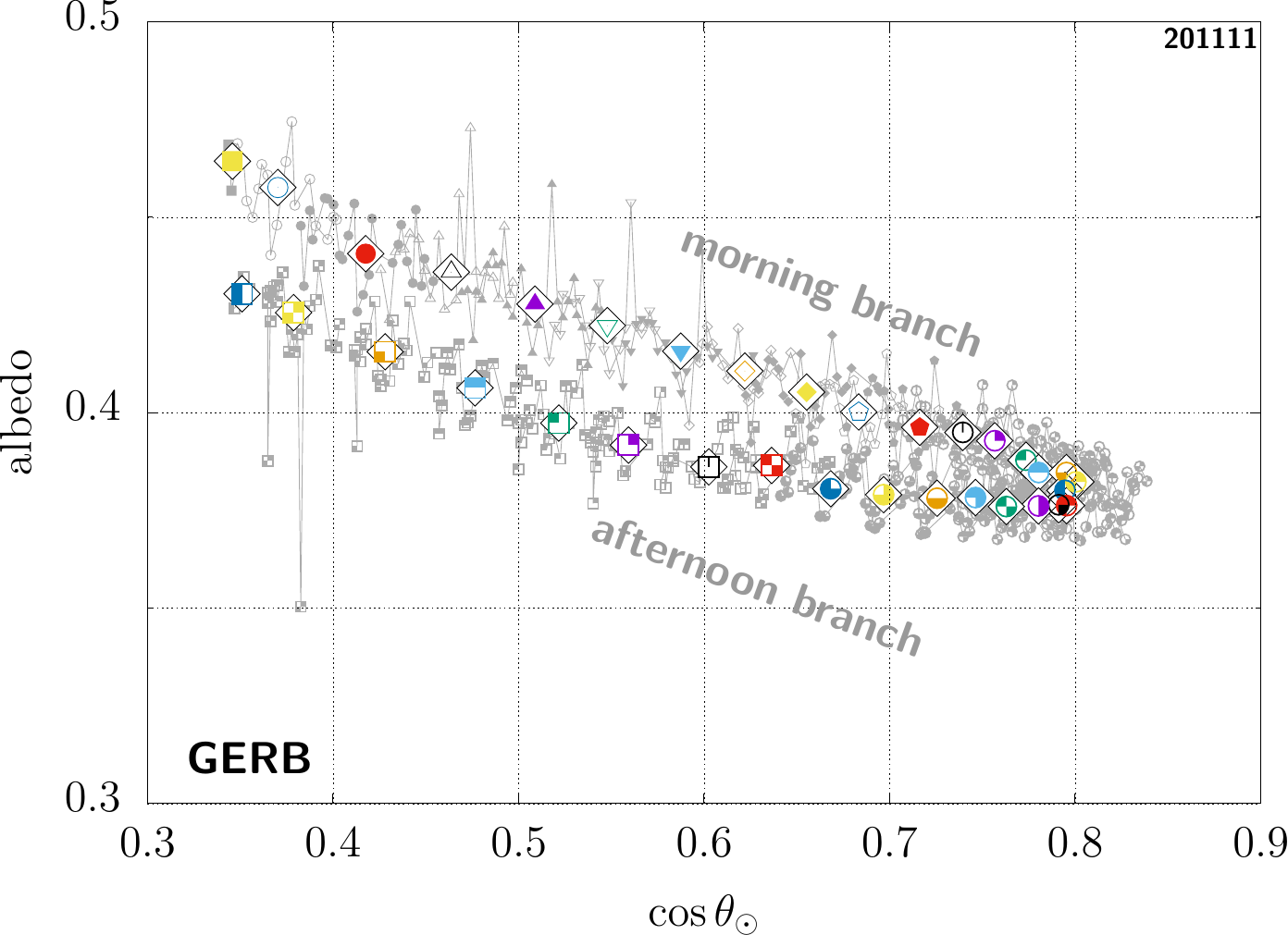}

\caption{Diurnal-asymmetry artefact in GERB data for the Sahara sample case shown in Figure~\ref{fig:samplepixel_sahara}. There have been at present no GERB Edition 1 data released for the period corresponding to the dual-view studied in this work (\textit{i.e.}\@ after 2016); this is then using GERB1 on \MSG{2} in November 2011. At the time, it was \MSG{2} which was located at a 0$\degree$ longitude, rather than \MSG{3} as in 2016, so this can be compared to the SEV3 case in Figure~\ref{fig:samplepixel_sahara} (\emph{left}); the exact same (colour, symbol) pairs are used here.}
\label{fig:samplepixel_sahara_GERB}
\end{figure}

A diurnal-asymmetry artefact is also clearly present when using pure GERB data.
This is shown in Figure~\ref{fig:samplepixel_sahara_GERB}, which presents results for the same Sahara sample GEO pixel first used in Section~\ref{sec:diurnal_asymmetry} to illustrate the existence of the diurnal-asymmetry artefact in GL-SEV data.
Note that, at the time of writing this paper, there have been no GERB Edition 1 data release covering the dual-view period studied here. For this reason, we had to use another year. Figure~\ref{fig:samplepixel_sahara_GERB} shows the results that we obtain with data from GERB1 (\MSG{2}), in November 2011, which was then at 0$\degree$ longitude. This can then be compared to our 2016 GL-SEV results from \MSG{3}; the albedo is similarly split on morning and afternoon branches with the morning branch being higher than the afternoon one in both cases.

This demonstrates that residual errors in the narrowband-to-broadband procedure alone would not be sufficient to explain the diurnal-asymmetry artefact discussed in Sections~\ref{sec:diurnal_asymmetry} and~\ref{sec:visualising_artefact}.
Indeed, GERB is a broadband instrument; it does not involve a narrowband-to-broadband procedure like the synthetic GL-SEV data.
Although we do not rule out a possible additional narrowband-to-broadband source of errors in the GL-SEV case, quantitatively this effect would likely be smaller or at most of the same order as what causes the artefact in GERB data shown in this appendix; definitely not much larger. 

\reftitle{References}
\externalbibliography{yes}
\bibliography{ourreferences}

\end{document}